\documentclass[12pt]{article}

\usepackage{scicite}
\usepackage{times}

\usepackage{xspace} 
\usepackage{gensymb}
\usepackage{caption}
\usepackage{subcaption}
\usepackage{graphicx}%
\usepackage{multirow}%
\usepackage{amsmath,amssymb,amsfonts}%
\usepackage{amsthm}%
\usepackage{mathrsfs}%
\usepackage[figuresright]{rotating}%
\usepackage[title]{appendix}%
\usepackage{xcolor}%
\usepackage{textcomp}%
\usepackage{manyfoot}%
\usepackage{booktabs}%
\usepackage{algorithm}%
\usepackage{algorithmicx}%
\usepackage{algpseudocode}%
\usepackage{program}%
\usepackage{listings}%
\usepackage{xr-hyper}
\usepackage{hyperref}
\usepackage[font=footnotesize]{caption}
\usepackage{setspace}

\newcommand{\ssftt}{SS~433\xspace}
\newcommand{\hess}{H.E.S.S.\xspace}
\newcommand{\jnoe}{HESS~J1908+063\xspace}
\newcommand{\xray}{X-ray\xspace}
\newcommand{\gammaray}{gamma-ray\xspace}

\newcommand{\thetitle}{Acceleration and transport of relativistic electrons in the parsec-scale jets of the microquasar SS~433}

\newenvironment{sciabstract}{%
\begin{quote} \bf}
{\end{quote}}

\title{\thetitle}

\author{H.E.S.S. Collaboration\footnote{\noindent laura.olivera-nieto@mpi-hd.mpg.de, brian.reville@mpi-hd.mpg.de, michelle.tsirou@desy.de, \newline
\indent\,\,\, jim.hinton@mpi-hd.mpg.de,  contact.hess@hess-experiment.eu}
\footnote{H.E.S.S. Collaboration authors and affiliations are listed in the supplementary materials}}

\date{}

\begin{document}
\newcommand*\aap{Astron. \& ~Astrophys.}
\let\astap=\aap
\newcommand*\aaps{Astron. \& ~Astrophys.~Suppl.}
\newcommand*\actaa{Acta Astron.}
\newcommand*\aj{Astron.~J.}
\newcommand*\ao{Appl.~Opt.}
\let\applopt\ao
\newcommand*\apj{Astrophys.~J.}
\newcommand*\apjl{Astrophys.~J.}
\let\apjlett\apjl
\let\apjsupp\apjs
\newcommand*\aplett{Astrophys.~Lett.}
\newcommand*\apspr{Astrophys.~Space~Phys.~Res.}
\newcommand*\araa{Annu. Rev. Astron. Astrophys}
\newcommand*\bac{Bull. astr. Inst. Czechosl.}
\newcommand*\bain{Bull.~Astron.~Inst.~Netherlands}
\newcommand*\caa{Chinese Astron. Astrophys.}
\newcommand*\cjaa{Chinese J. Astron. Astrophys.}
\newcommand*\fcp{Fund.~Cosmic~Phys.}
\newcommand*\gca{Geochim.~Cosmochim.~Acta}
\newcommand*\grl{Geophys.~Res.~Lett.}
\newcommand*\iaucirc{IAU~Circ.}
\newcommand*\icarus{Icarus}
\newcommand*\jcap{J. Cosmology Astropart. Phys.}
\newcommand*\jcp{J.~Chem.~Phys.}
\newcommand*\jgr{J.~Geophys.~Res.}
\newcommand*\jqsrt{J.~Quant.~Spectr.~Rad.~Transf.}
\newcommand*\memsai{Mem.~Soc.~Astron.~Italiana}
\newcommand*\mnras{Mon.~Not.~R.~Astron.~Soc.}
\newcommand*\nat{Nature}
\newcommand*\nphysa{Nucl.~Phys.~A}
\newcommand*\pasj{Publ.~Astron.~Soc.~Jpn}
\newcommand*\pasp{Publ.~Astron.~Soc.~Pac.}
\newcommand*\physrep{Phys.~Rep.}
\newcommand*\physscr{Phys.~Scr}
\newcommand*\planss{Planet.~Space~Sci.}
\newcommand*\pra{Phys.~Rev.~A}
\newcommand*\prb{Phys.~Rev.~B}
\newcommand*\prc{Phys.~Rev.~C}
\newcommand*\prd{Phys.~Rev.~D}
\newcommand*\pre{Phys.~Rev.~E}
\newcommand*\prl{Phys.~Rev.~Lett.}
\newcommand*\rmxaa{Rev. Mexicana Astron. Astrofis.}
\newcommand*\solphys{Sol.~Phys.}
\newcommand*\sovast{Soviet~Ast.}
\newcommand*\ssr{Space~Sci.~Rev.}

\maketitle

\begin{sciabstract}
	\ssftt is a microquasar, a stellar binary system with collimated relativistic jets. 
	We observed \ssftt in gamma rays using the High Energy Stereoscopic System (\hess), finding an energy-dependent shift in the apparent position of the gamma-ray emission of the parsec-scale jets. 
	These observations trace the energetic electron population and indicate the gamma rays are produced by inverse-Compton scattering. 
	Modelling of the energy-dependent gamma-ray morphology constrains the location of particle acceleration and requires an abrupt deceleration of the jet flow. 
	We infer the presence of shocks on either side of the binary system at distances of 25 to 30 parsecs and conclude that self-collimation of the precessing jets forms the shocks, which then efficiently accelerate electrons.
\end{sciabstract}

\begin{figure*}
	\centering
	\includegraphics[width=\textwidth]{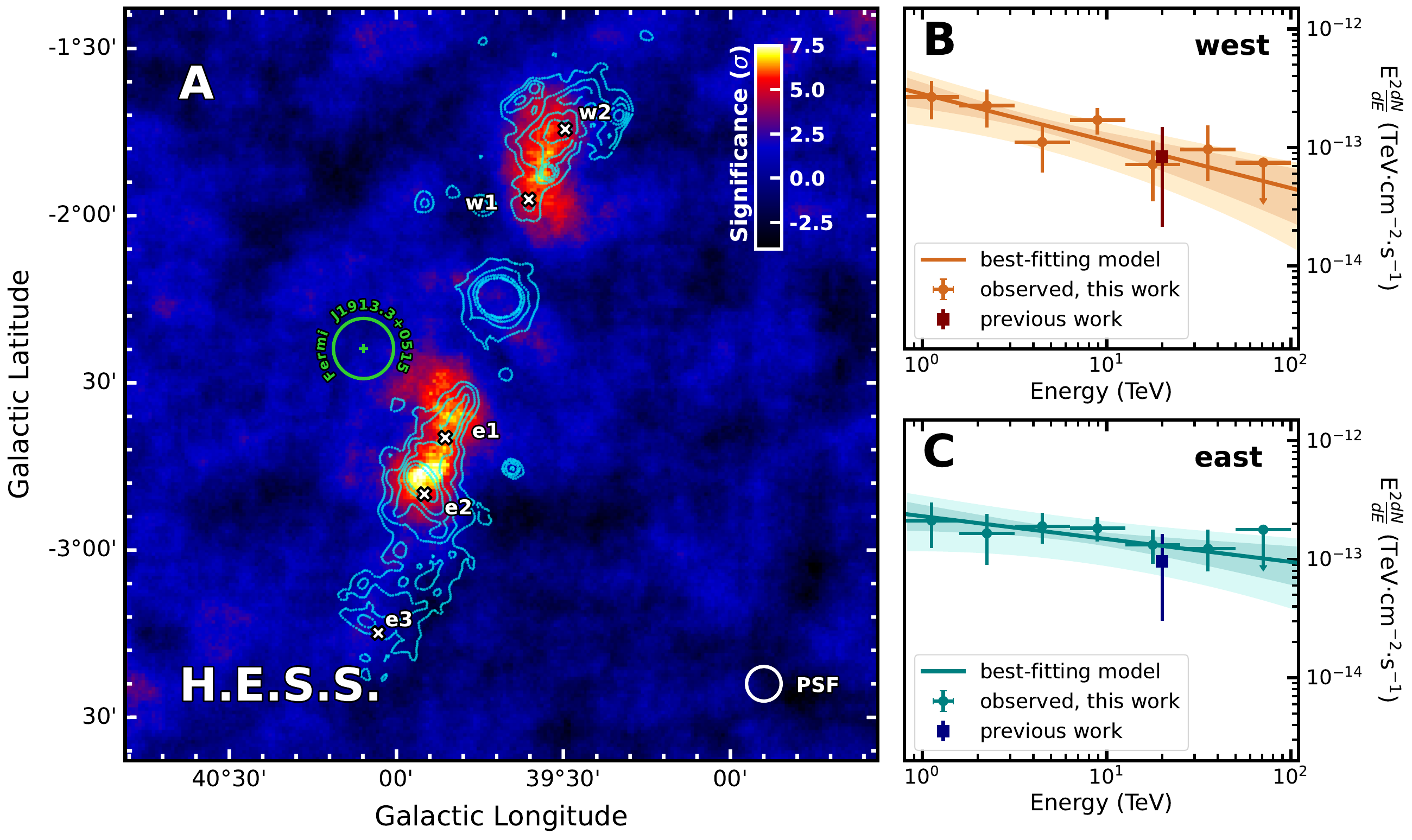}
	\caption{\textbf{Gamma-ray observations of \ssftt }. \textbf{A: }Significance map of the \hess observations at energies $>0.8$~TeV (color bar). 
		Cyan contours show the \xray emission~\cite{Brinkmann1996, Safi-Harb1997}. 
		White crosses indicate locations of \xray regions discussed in the text, w1, w2, e1, e2 and e3. 
		Significance is for the \hess excess counts above the background before accounting for statistical trials and after subtraction of the extended source \jnoe (subtraction shown in~Figure~\ref{fig:large_significance}). 
		The map has been smoothed with a top-hat function of radius 0.1\degree\xspace. 
		The white circle indicates the 68\% containment region of the \hess point-spread function (PSF). 
		The green cross indicates the position of Fermi~J1913+0515 and the green circle is its uncertainty. \textbf{B: }Orange circular points show our observed spectral energy distribution of the gamma-ray emission from the western jet. The brown square point is from previous observations~\cite{Abeysekara2018}. Error bars indicate the combined statistical (1$\sigma$) and systematic uncertainties; downward arrows indicate upper limits at 95\% confidence. 
		The solid line is the best-fitting power-law function, with dark and light shaded regions indicating the statistical and systematical uncertainties, respectively.
		\textbf{C: }Same as panel \textbf{B} but for the eastern jet. The regions from which the spectra shown in panels \textbf{B} and \textbf{C} were extracted are shown in~Figure~\ref{fig:large_significance}B. }
	\label{fig:significance}    
\end{figure*}

\newpage
\ssftt (V1343~Aql) is a binary system comprising a compact object, likely a black hole~\cite{Seifina2010, Cherepashchuk2019, Cherepashchuk2021}, and a type~A supergiant star~\cite{Hillwig2008}. 
Accretion onto the black hole causes it to launch a pair of jets moving in opposite directions at approximately one quarter of the speed of light $c$~\cite{Margon1979a, Eikenberry2001,Blundell2004}, almost perpendicular to our line of sight~\cite{Roberts2010}. The jets precess with a half-opening angle of 20\degree\xspace and a period of 162 days~\cite{Fabian1979, Milgrom1979, Margon1979, Fabrika2004}. 
Adopting the distance measurement of 5.5~kiloparsecs~(kpc)~\cite{Blundell2004}, optical and radio observations have shown that the precessing jets extend to distances of $\sim 10^{-3}$\,pc~\cite{Davidson1980} and $\sim 0.1$\,pc~\cite{Blundell2004} from the black hole, respectively. 
\xray emission reappears 25~pc from the binary (Figure~\ref{fig:significance}), indicating  collimated flows (the outer jets) on larger scales, which emit \xray photons via non-thermal processes~\cite{Brinkmann1996, Safi-Harb1997, Safi-Harb2022, Kayama2022}.\\

The outer jets terminate $\sim$100~pc from the black hole\cite{Brinkmann1996}, where they deform the surrounding radio nebula (known as W\,50 or SNR~G039.7-02.0) which is thought to be the supernova remnant associated with the formation of the compact object in \ssftt~\cite{Elston1987}. 
The morphology of W\,50 indicates that the opening angle of the outer jets is considerably smaller than the 20\degree\xspace precession angle of the inner jets~\cite{Goodall2011a}; the origin of this discrepancy is unknown~\cite{BowlerKeppens2018}. 
The lack of apparent change in the measured positions of radio filaments in the jet termination regions over a 33 year period provides an upper limit on their velocity of
$<0.023c$~\cite{Goodall2011, Sakemi2021}, though it is unclear whether the radio filaments trace the jets' flow. 
Bright \xray synchrotron knots have been observed in the outer jets but the temporal baseline and angular resolution was insufficient to determine their velocity. The dynamics of the outer jets and their termination process are poorly understood. \\

Since the initial \xray detection of the outer jets~\cite{Watson1983}, several attempts have been made to probe the non-thermal aspects and internal dynamics of the eastern~\cite{Brinkmann2007, Safi-Harb2022} and western~\cite{Kayama2022} outer jets.
However, observations of the \xray synchrotron emission alone cannot resolve variations in the distribution of accelerated particles. 
The intensity of synchrotron emission is approximately proportional to the number density of accelerated electrons and the energy density of the magnetic field; the latter is poorly constrained. 
\xray emitting electrons can also up-scatter low energy photons to the gamma-ray regime via the Inverse Compton scattering process.
This process directly traces the population of high-energy electrons, since the diffuse low-energy photon distribution in the Galaxy is expected to be smooth on the spatial scale of the outer jets~\cite{Popescu2017, SOM}.
Previous observations of TeV gamma-rays emitted by the outer jets of \ssftt~\cite{Abeysekara2018} indicate that the same energetic electrons responsible for the \xray emission also produce gamma rays via inverse Compton scattering~\cite{Bordas2010,Sudoh2020}. However, the angular resolution was insufficient to determine the emission regions and therefore the source of the energetic particles. \\

\subsection*{\hess observations of \ssftt}
We imaged the outer jets of \ssftt at TeV energies using the \hess array of imaging atmospheric Cherenkov telescopes. 
The observations totalled over 200 hours of exposure time. 
They were analysed using previously described methods that were optimised for faint sources and improved performance at the highest energies~\cite{OliveraNieto2022}. 
The extended source \jnoe (MGRO~J1908+06) contaminates part of the \ssftt jet, so was modelled then subtracted from the data (Figures~\ref{fig:large_significance} and~\ref{fig:profiles-jnoe}). The resulting gamma-ray image (Figure~\ref{fig:significance}A) shows two regions of gamma-ray emission at the known positions of the eastern and western jets, with peak statistical significances of 7.8$\sigma$ and 6.8$\sigma$, respectively. No significant ($>5\sigma$) emission is detected from the central binary or the eastern termination region (Figure~\ref{fig:significance}A), as we expect since the \xray emission from those regions is predominantly thermal~\cite{Marshall2002,Brinkmann1996}. Fermi~J1913+0515 (right ascension $=288.28°\pm0.04$\degree\xspace, declination $=5.27°\pm0.04$\degree\xspace) is a GeV gamma-ray source found to pulsate with a period consistent with the jet precession~\cite{Li2020}, suggesting a connection with the SS 433 system. No significant TeV emission is detected from this source~\cite{SOM}. Figure~1B-C shows the measured spectral energy distributions of each of the jets.
\\

\begin{figure*}
	\centering
	\includegraphics[width=1\textwidth]{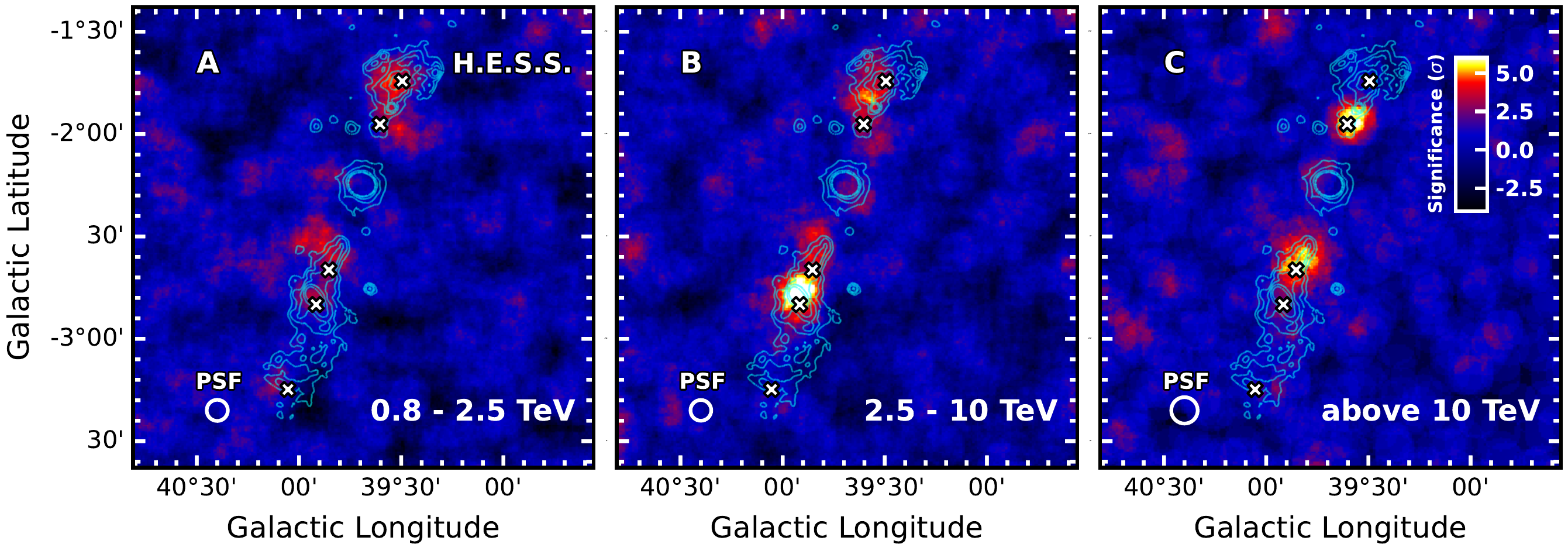}
	\caption{\textbf{Gamma-ray observations in different energy bands.} Same as Figure~\ref{fig:significance}A, but split into gamma-ray energy bands of (\textbf{A}) 0.8 to 2.5 TeV, (\textbf{B}) 2.5 to 10~TeV and (\textbf{C}) $>$10~TeV.}
	\label{fig:significance-energy}       
\end{figure*}

To investigate the energy-dependence of the gamma-ray emission, we split the full energy range into three bands (0.8 to 2.5, 2.5 to 10 and $>$10~TeV), which were selected to have approximately the same \gammaray excess counts over the background in each band. 
Figure~\ref{fig:significance-energy} shows the significance maps for each band. We detect significant ($>5\sigma$) gamma-ray emission along both jets for the two highest energy bands. 
In the lowest energy band we find lower-significance evidence of emission at 4.4 and 4.7$\sigma$ for the eastern and western jets, respectively. 
Gamma-ray emission $>$10~TeV appears only at the base of the outer jets visible in \xray for both the eastern and western jets. 
In contrast, lower energy gamma rays have their peak surface brightnesses at locations further along each jet, except for the lowest-energy band on the eastern side. 
In the latter case, no significant emission is detected inside the \xray jet region and evidence for emission appears close to the outer jet base (Figure~\ref{fig:significance-energy}A). 
In the western jet, the best-fit positions of the gamma-ray emission in each energy band have distances from the central binary (Table~\ref{tab:energy-dep}) that differ from each other by 0.97$\sigma$ and 2.6$\sigma$ when comparing adjacent energy bands, and by 5.3$\sigma$ when comparing the lowest and highest energy band. 
The equivalent values for the eastern jet are 2.6$\sigma$, 3.3$\sigma$ and 0.1$\sigma$. 
Our significance calculations include both systematic and statistical sources of uncertainty and a trials factor correction~\cite{SOM}. \\

\subsection*{Location of the particle acceleration}
We interpret the offsets between the emission in different energy bands as indicating that transport of particles in the outer jets is dominated by the bulk jet flow (advection) and not the random scattering of the particles on magnetic field fluctuations (diffusion).
The energy-dependent morphology then reflects an energy-dependent particle energy loss timescale. We infer that the emission arises from relativistic electrons, and not hadrons, because the loss timescale for hadronic processes depends only very weakly on particle energy~\cite{Aharonian2004}. The dominant energy loss mechanisms for high-energy electrons is likely to be synchrotron cooling. We conclude that the observed gamma-ray emission is the result of inverse Compton scattering~\cite{Blumenthal1970,Aharonian2004} of photons by high-energy electrons. 
Iron and other heavy nuclei are known to be present in the jet~\cite{Migliari2002}, so they might also be accelerated in the same region, but our observations cannot be used to constrain their presence (see Supplementary Text).\\

\begin{figure}[]
	\centering
	\includegraphics[width=0.95\textwidth]{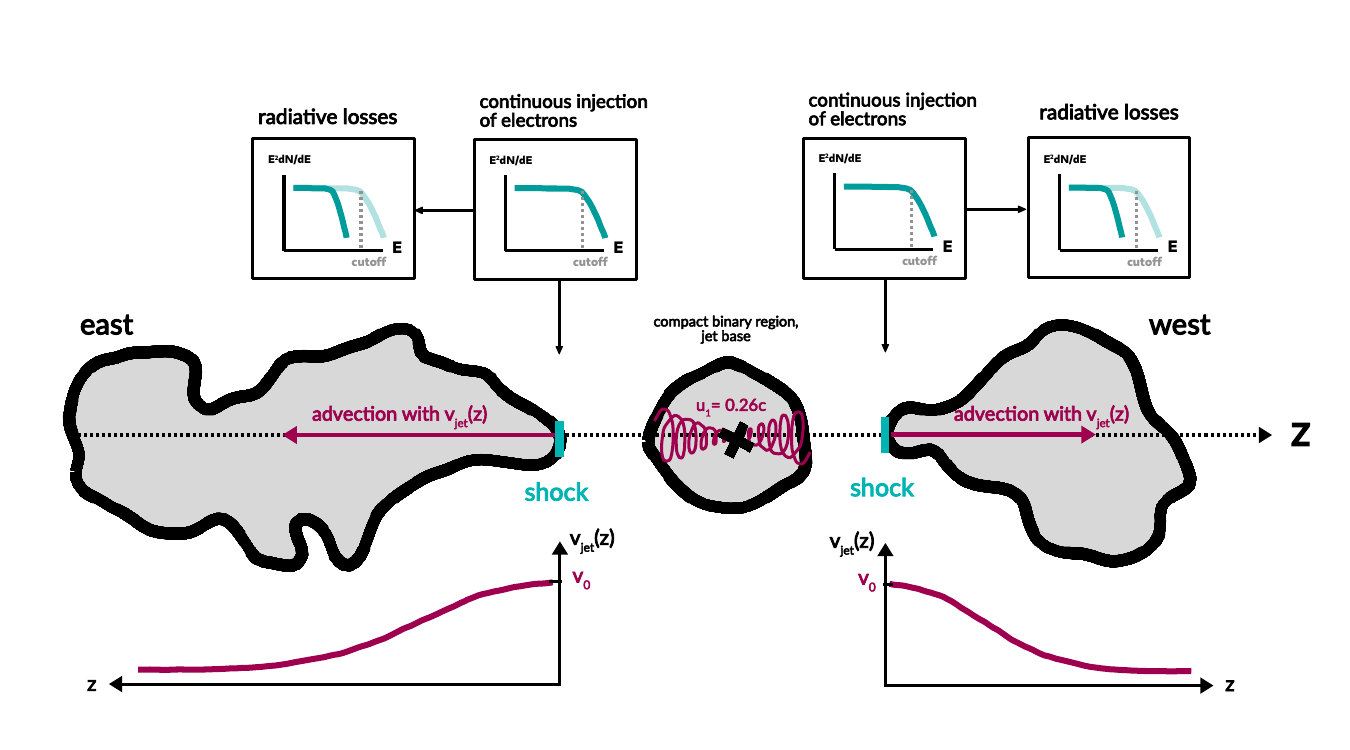}
	\caption{\textbf{Schematic diagram of our model. }
		Thick black lines roughly outline the \xray emission (grey shading) from the central region and the outer jets on the plane of the sky (rotated from the orientation in Figure~\ref{fig:significance}). 
		The precessing jet is launched with velocity $u_1\approx $ 0.26$c$ (purple spirals) and travels until it encounters a shock discontinuity (cyan bars), at the base of the outer jets. 
		Our 1D model injects electrons continuously at the outer jet base, with an energy spectrum derived from a fit to multi-wavelength observations from the outer jets (Table~\ref{tab:model-parameters}). 
		The injected electrons lose energy due to radiative losses, which affects their spectrum (indicated by the insets above the diagram). 
		Particles are transported along the jets by the combination of diffusion (not depicted) and advection along the jet axis coordinate $z$ with the jet flow at velocity $v_{\mathrm{jet}}(z)$ (purple arrows). 
		The velocity at the base of the jets behind the shock $v_0$ can be determined by fitting the model prediction to the \hess data. 
		We assume the jet flow decelerates as it moves away the jet base, indicated by the purple curves below the diagram. We also considered the alternative case of a constant velocity jet (see Supplementary Text).}
	\label{fig:sketch}
\end{figure}

The shorter cooling time of higher-energy electrons limits the distance from the acceleration site within which they can radiate, because they are transported away by either diffusion or advection. 
The absence of emission above 10~TeV at the location of the \xray knots (e2 and w2, Figure~\ref{fig:significance}A) indicates that they cannot be sites of particle acceleration to TeV energies, contradicting previous interpretations~\cite{Abeysekara2018, Kimura2020}. 
Instead, the concentration of emission above 10~TeV at the base of the \xray emission from the outer jets indicates this region is the site of particle acceleration to very high energies. 
We interpret the energy-dependent position of the gamma-ray emission in the jets of \ssftt as a consequence of the cooling and transport of particles that are accelerated at the base of the outer jets. 
Figure~\ref{fig:sketch} shows a schematic diagram of our proposed interpretation.\\

\subsection*{Modelling the outer jet dynamics}
Previous studies have shown that the jets are launched from the black hole with initial velocities of $u_1\approx $ 0.26$c$~\cite{Margon1979a, Eikenberry2001,Blundell2004}. 
We combined the distances between the gamma-ray excess regions in different energy bands with the electron cooling timescales~\cite{SOM} to determine the velocity $v_0$ of the outer jets at their base, approximately 25~pc away from the central binary. 
This calculation requires us to assume a spatial dependence of the deceleration of the jets as a function of the distance from the central binary. 
We used the observed opening angle of the jets in \xray images to determine the deceleration profile by assuming the jet flow is incompressible (Figure~\ref{fig:input_velocity_profile}). 
We also considered a jet propagating with constant velocity, under different energy loss assumptions, which leads to consistent values of $v_0$ (see Supplementary Text). 
Our observations cannot distinguish between the different jet propagation scenarios considered.\\

\sloppy
We model the energy-dependent morphology of the gamma-ray emission using a one-dimensional Monte Carlo simulation which includes radiation and cooling of particles as they are transported along the jet~\cite{SOM}. 
The model injects electrons at the base of the outer jet with an energy spectrum assumed to be of the form $dN/dE \propto E^{-\Gamma_{\rm e}}\exp(-\frac{E}{E_{\mathrm{cut}}})$ (Figure~\ref{fig:sketch}), where $N$ is the number of electrons, $E$ their energy and $\Gamma_{\rm e}$ and $E_{\mathrm{cut}}$ the spectral index and cutoff energy, respectively. 
We determine the best-fitting parameters of the injected electron spectrum and the average local magnetic field strength from the multi-wavelength spectral energy distribution of each outer jet separately. 
The value of $E_{\mathrm{cut}}$ is not constrained by the data; we find only a lower limit of $>200$~TeV at 68\% confidence level (C.L.). 
The model assumes the injection is continuous for 10\,000 years, this timescale being constrained by the combination of existing GeV gamma-ray flux upper limits~\cite{Fang2020} and the measured TeV gamma-ray flux (Figure~\ref{fig:age}). This electron injection timescale is consistent with previous dynamical estimates for the age of the W50/SS433 complex, which range between 10\,000 and 100\,000~yr~\cite{Goodall2011a,BowlerKeppens2018}. 
The simulation evolves the electron population numerically in discrete time steps. 
In each step, electrons are advected with the local jet velocity, then diffuse along the jet axis (neglecting transverse diffusion) and cool radiatively. 
This leads to an energy- and spatially-dependent electron distribution, from which we calculate one-dimensional profiles of the resulting non-thermal emission. 
We find that the resulting spatial distribution in the gamma-ray range only weakly depends on the parameters of the injected particle distribution~\cite{SOM}. \\

\begin{figure*}
	\centering
	\includegraphics[width=0.98\textwidth]{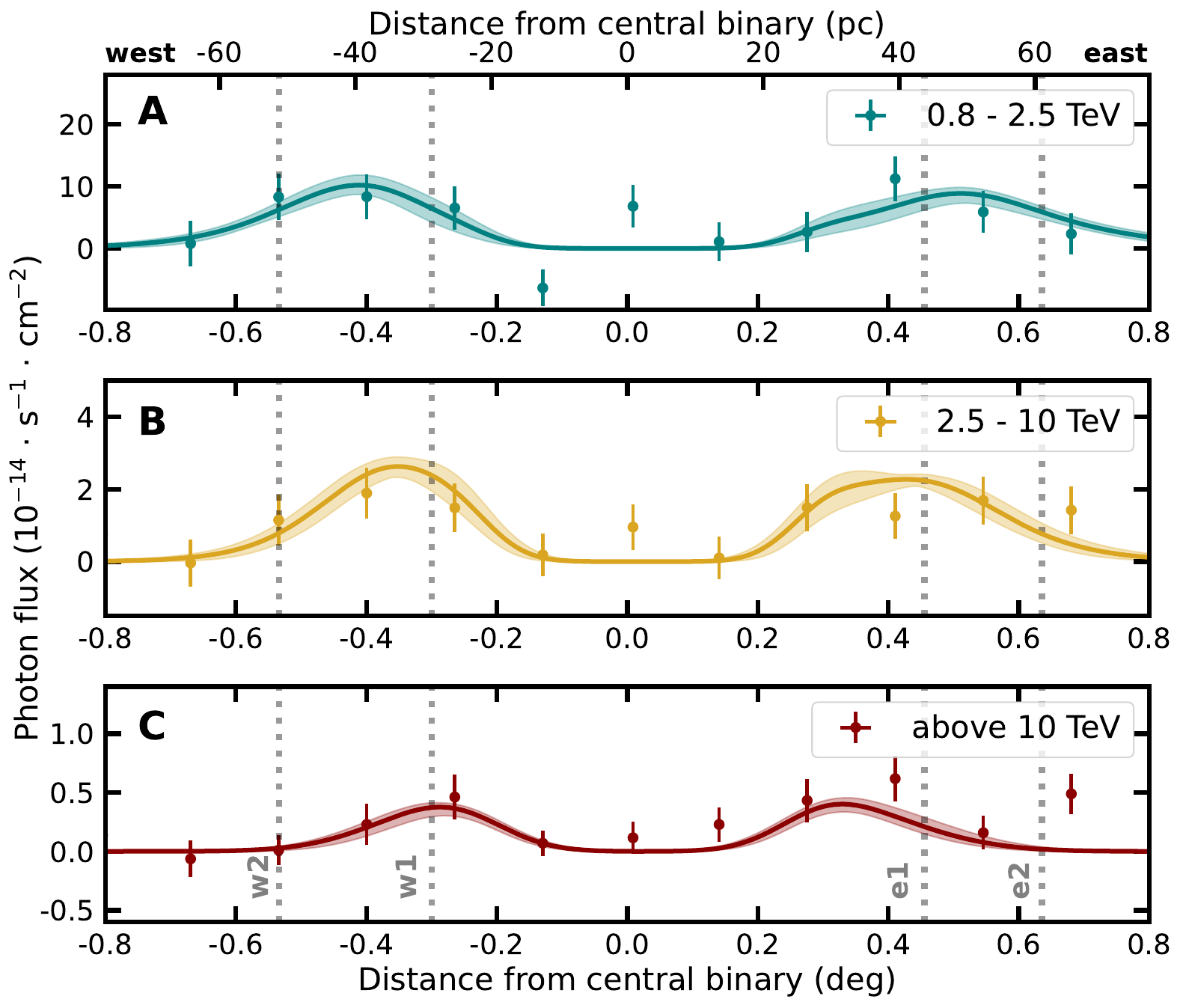}
	\caption{\textbf{Gamma-ray flux profiles along the jets compared with the model prediction.} Data points indicate the measured gamma-ray flux in spatial bins of 0.14\degree\xspace along the axis joining both jets through the central binary in the same three energy bands (panels \textbf{A}, \textbf{B} and \textbf{C}) as Figure~\ref{fig:significance-energy}. Error bars indicate the combined statistical (1$\sigma$) and systematic uncertainties.  
		Solid lines show the prediction of our best-fitting 1D model. 
		The shaded areas represent the combined statistical uncertainty of the best-fitting parameters. 
		Dashed grey vertical lines represent the positions of the \xray regions e1, e2, w1 and w2 (see Figure~\ref{fig:significance}), which are labelled in panel \textbf{C}. The top axis assumes a distance to the system of 5.5~kpc~\cite{Blundell2004}.}
	\label{fig:profile}       
\end{figure*}

Using the \hess data, we derive spatial profiles of the gamma-ray flux along the axis joining both outer jets through the central binary in the same three energy bands used in Figure~\ref{fig:significance-energy}. 
We fitted the resulting model emission profile to the data with $v_0$ and the diffusion coefficient, the latter assumed to be spatially uniform, as free parameters. 
The injected electron spectrum parameters are fixed to the values obtained from the fit to the multi-wavelength spectral energy distributions described above.
We assume the same starting velocity for both the eastern and western jet. 
The best-fitting value is $v_0=(0.083\pm0.026_{\mathrm{stat}}\pm0.010_{\mathrm{syst}})c$. The systematic uncertainty is derived from the choice of parameters for the injected electron spectrum~\cite{SOM}. Figure~\ref{fig:profile} shows the gamma-ray spatial profiles and the best-fitting model.\\

\subsection*{Interpretation as a standing shock}
Our modelling shows the data are consistent with the presence of a particle accelerator, likely a shock, at the base of the \ssftt outer jets which is capable of accelerating particles to very high energies. 
Our lower limit on $E_{\mathrm{cut}}$ indicates the acceleration of electrons to energies $>$200~TeV (68\% C.L.). 
At the inferred magnetic field strength of approximately 20~\textmu G~(Table~\ref{tab:model-parameters}), to keep up with cooling the acceleration rate must be close to the theoretical maximum, assuming diffusive shock acceleration~\cite{Drury1983, SOM}. 
Therefore the jet flow cannot have decelerated much from its inferred launch velocity of 0.26$c$ prior to reaching the shock, because if it had, the particle acceleration could not compete with radiative losses at electron energies above several hundred~TeV (Figure~\ref{fig:cooling}). 
The velocity we infer at the base of the outer jets $v_0$ is a fraction $\chi=0.319\pm 0.10_{\mathrm{stat}}\pm0.039_{\mathrm{syst}}$ of the jet launch velocity. 
This is compatible with the velocity ratio expected for a sub-relativistic shock, which is $\chi\approx 0.25$~\cite{Landau1987, Kirk1999}. 
A shock at this location is consistent with the spatial coincidence between the position of the highest-energy gamma-ray emission and the location of the recently reported sharp \xray reappearance of the \xray emission~\cite{Safi-Harb2022,Kayama2022}. 
This region has previously been interpreted as the acceleration site, but without involving shocks~\cite{Safi-Harb2022}. 
Here we have shown that if the advection in the jet flow is taken into account, the observations are consistent with the shock acceleration scenario.
Our observations also constrain the velocity of the shock, which would have needed to advance a small distance ($\ll 10$~pc) in the lifetime of the TeV gamma-ray emitting electrons (Figure~\ref{fig:cooling}).\\

There is no single model that has yet reproduced all the observational features of \ssftt~\cite{Goodall2011a,Ohmura2021}. 
While simulations can account for the observed difference in opening angle between the inner and outer jets due to the action of the ambient medium~\cite{MonceauBaroux2014, MonceauBaroux2015,BowlerKeppens2018}, this process would take place near the binary and does not result in the observed sharp transitions or shocks at 25 to 30~pc. 
The mirrored reappearance of the jets at this distance implies a physical significance to this radius, though there is no further observational evidence to indicate that this location is special.
Radio observations of the jet launch region reveal the presence of an equatorial outflow perpendicular to the axis of the jets~\cite{Blundell2001}.
Such an outflow has previously been argued to result in the formation of a quasi-spherical shock at distances of tens of pc from the binary~\cite{Konigl1983}. 
However, the \xray shell that would be produced by such a shock has not yet been detected.\\

The proximity of \ssftt to Earth allows us to investigate shock physics and associated non-thermal processes in mildly relativistic jets. 
These insights can be applied both to other microquasars~\cite{Kantzas2021} as well as to the larger and more distant jets launched from the centres of other galaxies, in which jet sub-structure cannot be resolved at high energies~\cite{Collaboration2020}.
Our results imply that shocks forming within jets accelerate particles at close to the theoretical maximum energy~\cite{Hillas1984,Aharonian2004}. 
Thus microquasars could be major contributors to the measured Galactic cosmic-ray flux at PeV energies, while extra-galactic jets could reach the EeV regime of ultra-high-energy cosmic rays (see Supplementary Text).

\newpage
\bibliographystyle{science}

\section*{Acknowledgements}

\noindent {\bf Acknowledgements} 
We thank Jian Li and Ke Fang for providing the reduced Fermi-LAT dataset and GeV flux points, respectively.
We appreciate the excellent work of the technical
support staff in Berlin, Zeuthen, Heidelberg, Palaiseau, Paris,
Saclay, Tübingen and in Namibia in the construction and operation
of the equipment. This work benefited from services provided by the
H.E.S.S. Virtual Organisation, supported by the national resource
providers of the EGI Federation.

\noindent {\bf Funding}
The support of the Namibian authorities and of the University of
Namibia in facilitating the construction and operation of H.E.S.S.
is gratefully acknowledged, as is the support by the German
Ministry for Education and Research (BMBF), the Max Planck Society,
the German Research Foundation (DFG), the Helmholtz Association,
the Alexander von Humboldt Foundation, the French Ministry of
Higher Education, Research and Innovation, the Centre National de
la Recherche Scientifique (CNRS/IN2P3 and CNRS/INSU), the
Commissariat à l’énergie atomique et aux énergies alternatives
(CEA), the U.K. Science and Technology Facilities Council (STFC),
the Irish Research Council (IRC) and the Science Foundation Ireland
(SFI), the Knut and Alice Wallenberg Foundation, the Polish
Ministry of Education and Science, agreement no. 2021/WK/06, the
South African Department of Science and Technology and National
Research Foundation, the University of Namibia, the National
Commission on Research, Science \& Technology of Namibia (NCRST),
the Austrian Federal Ministry of Education, Science and Research
and the Austrian Science Fund (FWF), the Australian Research
Council (ARC), the Japan Society for the Promotion of Science, the
University of Amsterdam and the Science Committee of Armenia grant
21AG-1C085. 

\noindent {\bf Author Contributions:}
L.~Olivera-Nieto performed the  \hess analysis and M.~Tsirou performed the cross-check analysis. N.~Tsuji analysed the \xray data. B.~Reville, L.~Olivera-Nieto and J.~Hinton performed the interpretation and modelling. The manuscript was prepared by L.~Olivera-Nieto, B.~Reville and M.~Tsirou. S. Wagner is the collaboration spokesperson. All other H.E.S.S. collaboration authors contributed to the design, construction and operation of H.E.S.S., the development and maintenance of data handling, data reduction or data analysis software. All authors meet the journal’s authorship criteria and have reviewed, discussed, and commented on the results and the manuscript.

\noindent {\bf Competing Interests:} The authors declare that they have no competing interests.

\noindent {\bf Data and Materials Availability:} The H.E.S.S. data are available at: \url{https://www.mpi-hd.mpg.de/HESS/pages/publications/auxiliary/2023_SS433/index.html}. This includes the counts and background maps, the subsequently derived significance and flux maps (Figures~1A and 2), the flux profiles (Figure~3) and the raw data and derived flux points from the spectral measurements of the jets (Figure~1B-C). Our modelling code is available at \url{https://github.com/LauraOlivera/particle-transport-1D}~\cite{OliveraNieto2023}.

\textbf{Supplementary Material: Authors and affiliations, Materials and Methods, Supplementary Text, Figure~S1-S16, Tables S1-S5, References (50-88)}

\newpage

\baselineskip24pt

\begin{center}
	\includegraphics[width=0.6\textwidth]{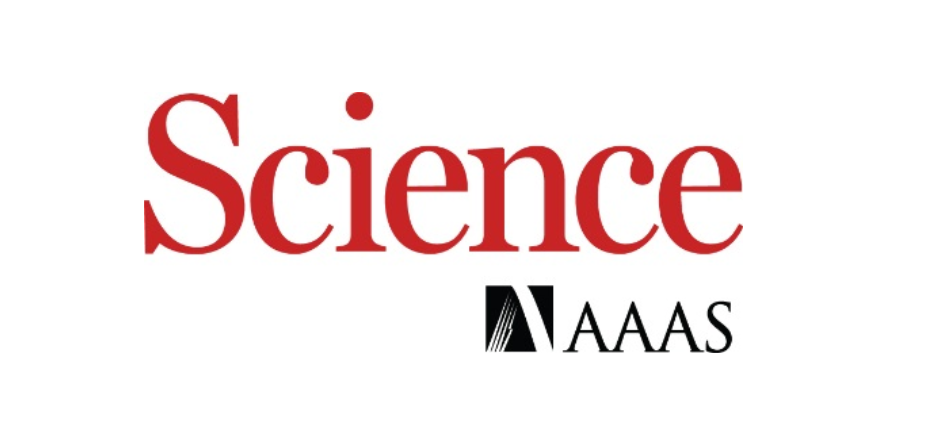}
\end{center}

\bigskip
\begin{center}
	{\LARGE Supplementary Material for}\\
	\bigskip
	{\Large \thetitle}\\
	\bigskip
	{\Large H.E.S.S. Collaboration*}\\
	\medskip
	{\par *Correspondence to: contact.hess@hess-experiment.eu; Laura Olivera-Nieto (laura.olivera-nieto@mpi-hd.mpg.de), Brian Reville (brian.reville@mpi-hd.mpg.de), Jim Hinton (jim.hinton@mpi-hd.mpg.de), Michelle Tsirou (michelle.tsirou@desy.de)}
\end{center}

\noindent
{\large This PDF file includes:}

\bigskip

\begin{spacing}{1.0}
	\noindent
	\hspace*{1.2cm} Authors and affiliations\\
	\hspace*{1.2cm} Materials and Methods\\
	\hspace*{1.2cm} Supplementary Text \\
	\hspace*{1.2cm} Figs. S1 to S16\\
	\hspace*{1.2cm} Tables S1 to S5
\end{spacing}

\setcounter{figure}{0}
\renewcommand{\thefigure}{S\arabic{figure}}
\setcounter{table}{0}
\renewcommand{\thetable}{S\arabic{table}}
\setcounter{equation}{0}
\renewcommand{\theequation}{S\arabic{equation}}

\newpage
\subsection*{H.E.S.S. Collaboration authors and affiliations}
\author{
	F.~Aharonian$^{1,2}$,
	F.~Ait~Benkhali$^{3}$,
	J.~Aschersleben$^{4}$,
	H.~Ashkar$^{5}$,
	M.~Backes$^{6,7}$,
	V.~Barbosa~Martins$^{8}$,
	R.~Batzofin$^{9}$,
	Y.~Becherini$^{10,11}$,
	D.~Berge$^{8,12}$
% ...
	\\
	$^{1}$Dublin Institute for Advanced Studies, Dublin D02 XF86, Ireland \\
	$^{2}$Max-Planck-Institut f\"ur Kernphysik, Heidelberg D-69117, Germany \\
	$^{3}$Landessternwarte, Universit\"at Heidelberg, Heidelberg D-69117, Germany \\
	$^{4}$Kapteyn Astronomical Institute, University of Groningen, Groningen 9747 AD, The Netherlands \\
	$^{5}$Laboratoire Leprince-Ringuet, École Polytechnique, Centre national de la recherche scientifique (CNRS), Institut Polytechnique de Paris, Palaiseau F-91128, France \\
	$^{6}$Department of Physics, University of Namibia, Windhoek 10005, Namibia \\
	$^{7}$Centre for Space Research, North-West University, Potchefstroom 2520, South Africa \\
	$^{8}$Deutsches Elektronen-Synchrotron (DESY), Zeuthen D-15738, Germany \\
	$^{9}$Institut f\"ur Physik und Astronomie, Universit\"at Potsdam, Potsdam 14476, Germany \\
	$^{10}$Laboratoire Astroparticule et Cosmologie, Université de Paris, CNRS, Paris F-75013, France \\
	$^{11}$Department of Physics and Electrical Engineering, Linnaeus University, V\"axj\"o 351 95, Sweden \\
	$^{12}$Institut f\"ur Physik, Humboldt-Universit\"at zu Berlin, Berlin D-12489, Germany \\
% ...
}
\makeatletter
\@author
\newpage

\section*{Materials and Methods}
\subsection*{The \hess array}
The \hess array of imaging atmospheric Cherenkov telescopes is located in the Khomas Highland of Namibia, at an altitude of 1,835 m. \hess is sensitive to gamma rays ranging from tens of GeV to tens of TeV. The array consists of five Cherenkov telescopes: four with mirror diameters of 12 m placed in a square configuration (designated CT1 to CT4) and a single telescope at the centre (CT5) with a mirror diameter of 28 m. The 12-m telescope array is sensitive to gamma rays of energies above several hundreds of GeV, while the central, large telescope is able to detect fainter Cherenkov emission, which in turn translates to lower gamma-ray energies~\cite{Aharonian2006}.

\subsection*{Observations of \ssftt}
The \ssftt region was observed by the \hess array of telescopes as part of a Galactic plane survey~\cite{HESSCollaboration2018}, and again during two dedicated campaigns. 
The first dedicated campaign, between 2009 and 2011 did not detect \ssftt and has been described previously~\cite{HESSMAGIC2018}. 
A second campaign, between 2018 and 2021, collected around 150~h of data. We combined all three datasets, resulting in around 200~h of data. Table~\ref{tab:data} lists the properties of each dataset we used. 
Some of this data was taken before an upgrade of the CT1-4 cameras~\cite{Ashton2020} which we refer to as the HESS-I dataset, and the data taken after the upgrade is labelled HESS-IU. 
For HESS-IU, we make a further distinction between the observations taken with (HESS-IU-CT5) and without (HESS-IU) CT5. 
All the observations involving CT5 were performed after an upgrade of the CT5 camera in 2019~\cite{Bi2022}. 
The HESS-IU and HESS-IU-CT5 datasets overlap in time. 
Data-taking was split into observation runs usually spanning 28 minutes. 
The pointing positions of the observations comprising the HESS-I dataset are concentrated around \jnoe and the location of the western jet of \ssftt. 
The pointing positions of the later observations were chosen to achieve a uniform exposure over the entire field.

\begin{table}[h]
	\centering
	\caption{\textbf{\hess datasets.} Listed are the exposure time, mean zenith angle, start and end dates and array configuration for each of the used datasets. The exposure quoted corresponds to the maximum value of the livetime when the exposure from the different pointing positions are combined in the field of view.}
	\begin{tabular}{c | c c c c c}
		dataset & exposure & mean zenith  & start date & end date & CT5 included  \\
		& (h) & (deg) &  & \\
		\hline
		HESS-I  & 71.1 & 38.8 & 2005 Jun 3 & 2013 Aug 20 & no\\
		HESS-IU  & 33.5 & 44.2 & 2018 Aug 29 & 2021 Nov 3 & no\\
		HESS-IU-CT5  & 111.1 & 48.9 & 2020 Jun 19 & 2021 Nov 5 & yes\\
		
	\end{tabular}
	
	\label{tab:data}
\end{table}

\subsection*{Event reconstruction and background rejection}
\hess records stereoscopic images of atmospheric showers produced by high-energy particles as they travel in the Earth's atmosphere. Gamma-ray-like events are selected by a Boosted Decision Tree classifier~\cite{Ohm2009}. 
The selected events are reconstructed using the Image Pixel-wise fit for Atmospheric Cherenkov Telescopes (ImPACT) algorithm~\cite{Parsons2014}, which uses a maximum-likelihood framework to fit a library of simulated templates to the data images. 
This process results in an estimate for the gamma-ray energy and direction, among other parameters. 
We only use information from CT1-4 for these steps. 
An extra step of background rejection is applied to the observation runs that include the CT5 telescope, which is the majority of runs taken in the 2019 to 2021 campaign. 
This method exploits the reduced threshold of the central telescope, which is more efficient at detecting faint emission from other particles, such as muons. 
Muons are produced in large numbers in hadronic showers, the main source of background for Cherenkov gamma-ray detectors~\cite{OliveraNieto2021}. 
In this step, the CT5 image of events selected as gamma-ray-like based on their CT1-4 image is compared to the associated ImPACT prediction image of that event. 
Events for which the CT5 image differs according to a set of criteria from the expected template are rejected, which leads to a factor 3 to 4 improvement in background rejection. 
The method criteria, performance and implementation have been described elsewhere~\cite{OliveraNieto2022}. 

\subsection*{Data reduction}
Subsequent steps in the data reduction and analysis were carried out using the gammapy software~\cite{Donath2023}, version 0.19~\cite{Donath2021}. 
The selected gamma-ray-like events were binned into a three-dimensional data cube, consisting of sky-map of 6\degree~width and 0.01\degree~bin size, centred at the position of \ssftt and an energy axis with 22 bins equally spaced in logarithmic energy between 0.63 and 100 TeV. 
For each observation, a safe energy range was derived, defined as the range in which the energy bias was less than 10\%~\cite{Aharonian2006}. 
This criterion rejected the energy bin between 0.63 and 0.8~TeV in all our observations, so 0.8~TeV is the lowest event energy in our \hess data.

\sloppy
Each observation run has an associated model of expected background counts which depends on radial offset from the pointing position and on reconstructed energy~\cite{Mohrmann2019}. 
These background models were derived from observation runs with mostly extra-galactic pointing positions in which expected gamma-ray sources were masked. 
The background model varies between runs due to different pointing altitude position and hardware changes. 
For each individual run in the dataset, the counts predicted by the model outside of an exclusion mask are fitted to those measured in the same region using two free parameters that determine the overall background normalisation and spectral shape. 
This procedure corrects the background model for possible variations due to atmospheric conditions and instrumental degradation~\cite{Mohrmann2019}. 
The exclusion mask was defined to cover known and expected gamma-ray sources in the \ssftt field of view. It is composed of a band of 2\degree~height around the Galactic plane ($\lvert b\rvert\leq 1\degree$ where $b$ is the Galactic latitude), a circle of 1.33\degree~radius for the nearby extended source \jnoe (Figure~\ref{fig:large_significance}) and two circles of 0.57\degree~radius for the jets of \ssftt.  

The instrument response functions (IRFs) that describe the precision of the energy and direction reconstruction, as well as the effective area of the detector were projected into multi-dimensional sky-maps with the corresponding energy axes. 
The set of maps corresponding to each observation were stacked by adding the counts and background maps and combining the IRFs weighted by the exposure of each run, considering only the energy range determined as safe for each observation. 
Significance maps were computed using a maximum-likelihood ratio test based on the comparison of the number of measured counts to the expected background counts~\cite{Li1983}. 
Both the background and counts map were smoothed with a top-hat kernel of radius 0.1\degree, chosen to be roughly the size of the \hess point-spread-function (PSF).
All data reduction was confirmed by an independent reduction, acting as a cross check. 
The cross-check analysis employed independent calibration, reconstruction and background suppression~\cite{Naurois2009}.

\begin{figure*}[]
	\centering
	\includegraphics[width=0.5\textwidth]{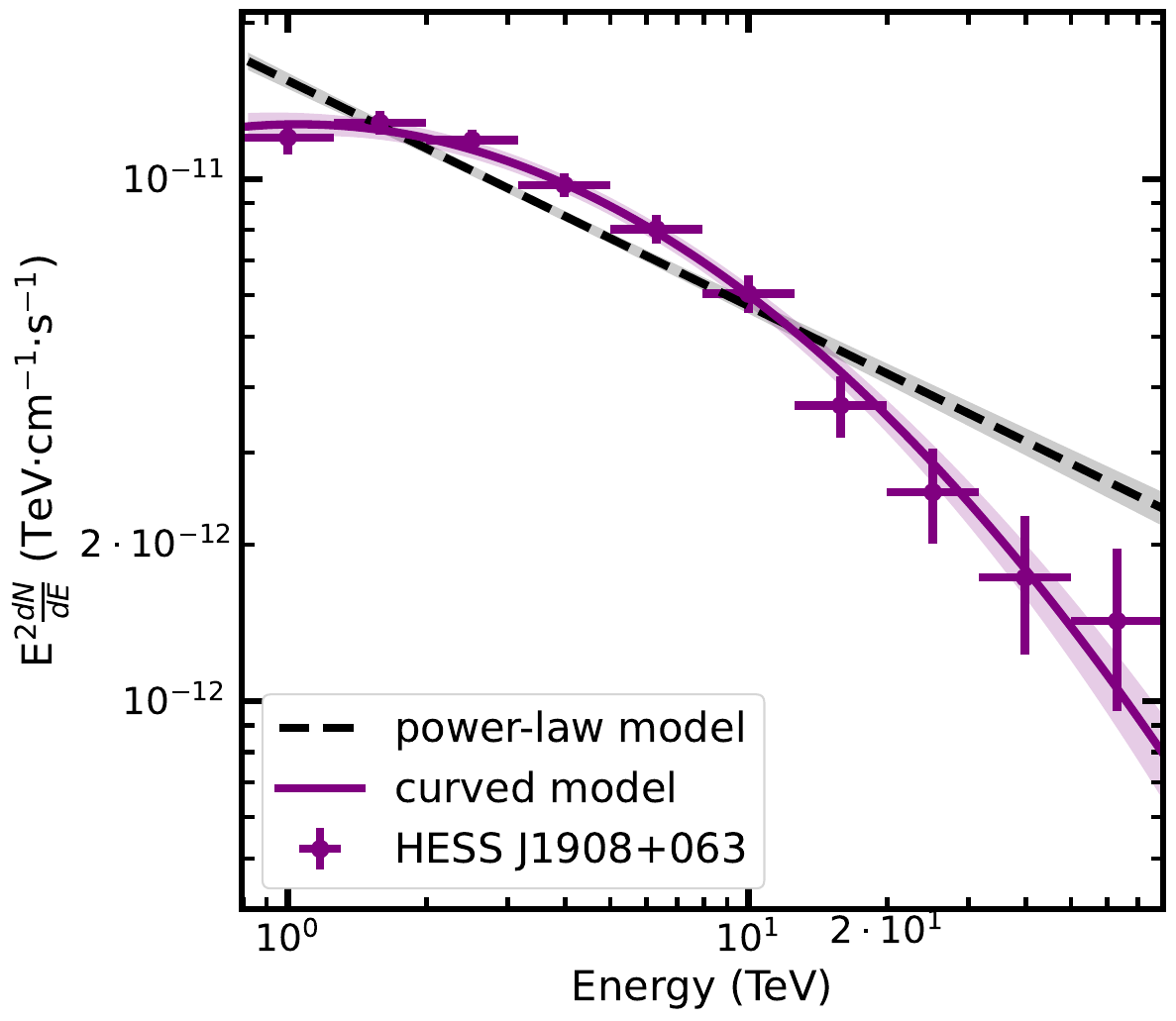}
	\caption{\textbf{Spectral energy distribution of \jnoe. } Purple circular points show our observed spectral energy distribution of the gamma-ray emission from \jnoe. Error bars indicate statistical (1$\sigma$) uncertainties. The solid purple line is the best-fitting log-parabola function. The dashed black line is the best-fitting power-law function.}
	\label{fig:jnoe-spec}
\end{figure*}

\subsection*{Removal of contaminating source \jnoe}
\begin{figure*}[]
	\centering
	\includegraphics[width=0.98\textwidth]{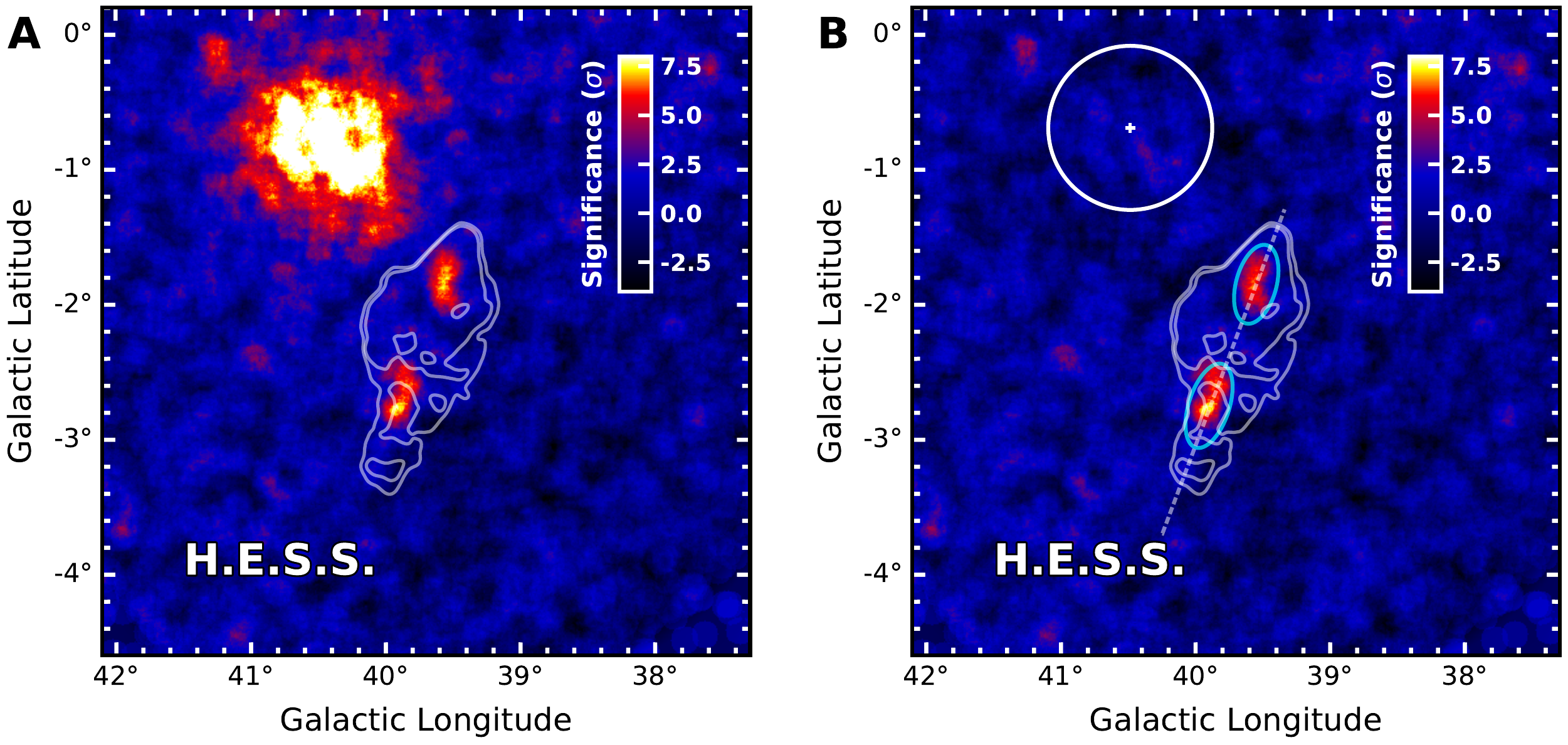}
	
	\caption{\textbf{Subtraction of \jnoe. }Same as Figure~\ref{fig:significance}A, but (\textbf{A}) before and (\textbf{B}) after subtracting the emission from the nearby extended source \jnoe. 
		In panel \textbf{B}, the white circle indicates the 68\% containment region of the model fitted to \jnoe, and the white cross is its bes-fitting position. 
		In both panels, the solid white contours show radio emission from the W\,50 nebula~~\cite{Reich1984,Reich1990, Furst1990}.
		In panel \textbf{B}, the blue ellipses show the regions from where the spectral measurement of the jets is extracted (Figure~\ref{fig:significance}B-C). The dashed line shows the axis across the jets used to derive the gamma-ray spatial profiles shown in Figures~\ref{fig:profile} and~\ref{fig:profiles-jnoe}.}
	\label{fig:large_significance}
\end{figure*}

The bright extended source \jnoe is located less than 2\degree\xspace away from the position of \ssftt and detected with significance of more than 10$\sigma$ in the combined dataset. 
To assess the degree of contamination into the \ssftt region, we fitted the emission from \jnoe with a combined spatial and spectral model using a maximum-likelihood framework in gammapy. 
The model of \jnoe has a single Gaussian component with curved spectrum described by a log-parabola function (Figure~\ref{fig:jnoe-spec}). 
The curvature is preferred to a simpler power-law at a significance of 6.4$\sigma$. Figure~\ref{fig:large_significance} shows the significance map of the full field of view, including the \jnoe region, before and after the best-fitting model is subtracted.

\begin{figure*}[]
	\centering
	\includegraphics[width=0.8\textwidth]{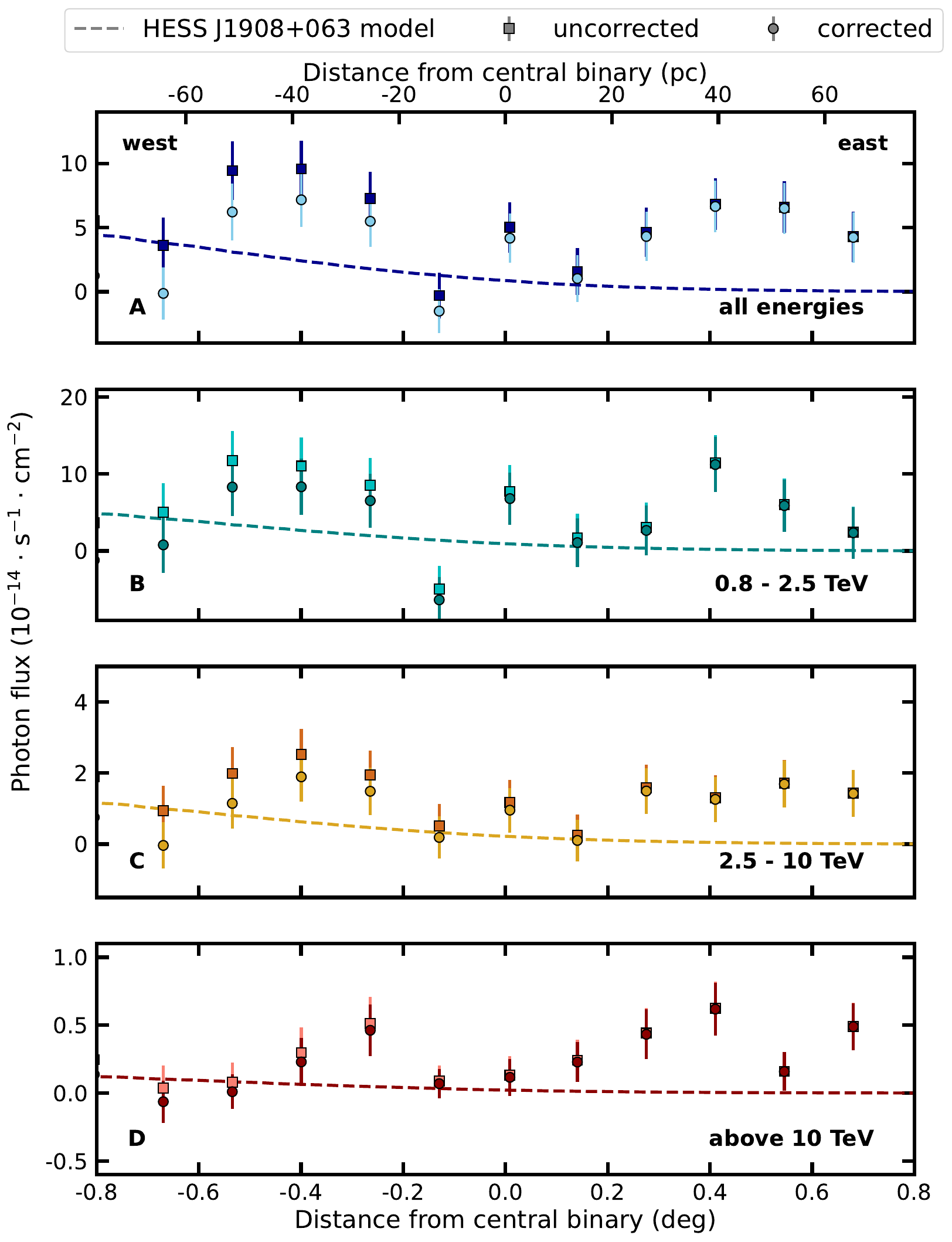}
	
	\caption{\textbf{Gamma-ray flux profiles along the jets showing the contamination of \jnoe.} The data points represent the measured flux in spatial bins of 0.14\degree\xspace along the axis joining both jets through the central binary (Figure~\ref{fig:large_significance}) for energies (\textbf{A}) above 0.8~TeV, (\textbf{B}) 0.8 to 2.5~TeV, (\textbf{C}) 2.5 to 10~TeV and (\textbf{D}) above 10~TeV. 
		Squares and circles indicate the flux before and after subtracting \jnoe. Error bars indicate the combined statistical (1$\sigma$) and systematic uncertainties. Circles in panels B-D are the same data as shown in Figure~\ref{fig:profile}. 
		The dashed lines show the flux of the \jnoe model at each location. The top axis assumes a distance to the system of 5.5~kpc~\cite{Blundell2004}.}
	\label{fig:profiles-jnoe}
\end{figure*}

Figure~\ref{fig:profiles-jnoe} shows the gamma-ray flux profiles measured along the jets of \ssftt for different energy ranges, before and after the subtraction of \jnoe. 
The contamination of \jnoe to the jets region is strongest at the tip of the western jet, located at around 0.5\degree\xspace from the binary. 
At that location, \jnoe contributes about $\approx40\%$ of the measured flux. 
This contribution quickly decreases towards the eastern jet, where it is negligible.
We include the subtraction of \jnoe in our estimation of systematic errors (see below).

\subsection*{Background systematic effects}
\begin{figure*}
	\centering
	\includegraphics[width=0.5\textwidth]{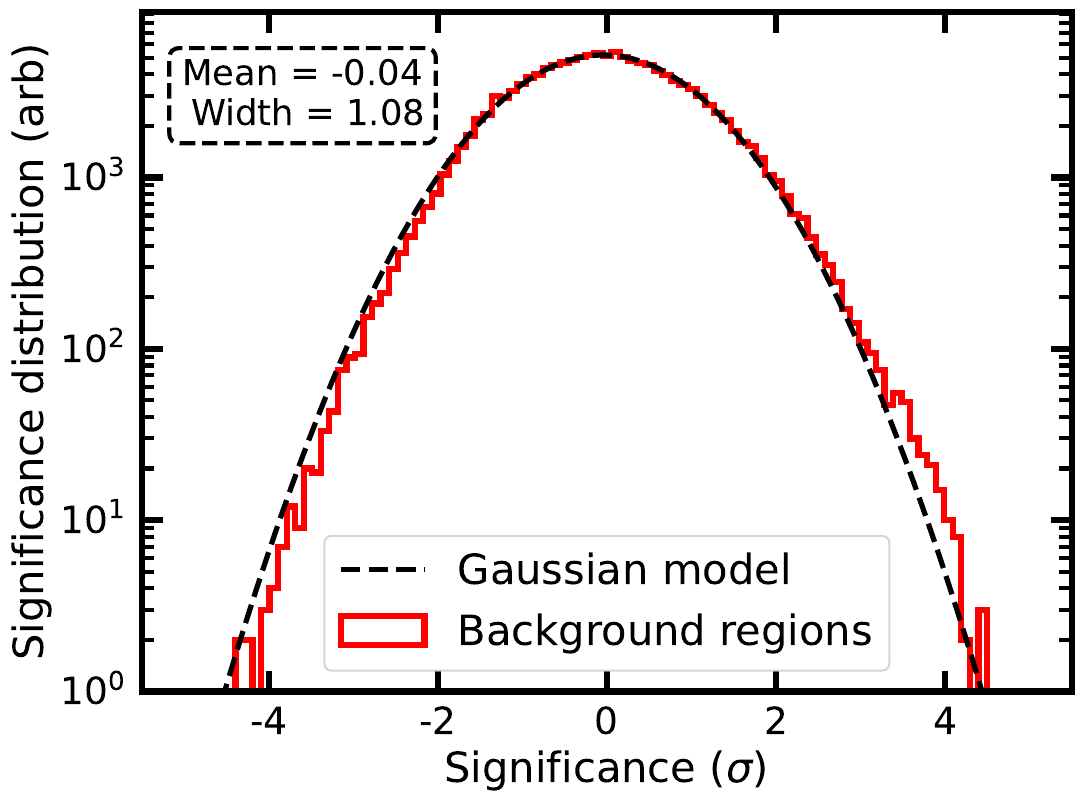}
	\caption{\textbf{Background significance distribution.} The distribution of significance values (red histogram) is plotted for all locations outside of an exclusion mask covering all expected sources. 
		The black dashed line is a Gaussian function fitted to the distribution (black line), Its mean and width are labelled. The deviation of this model from the expected mean of 0 and width of 1 is used to assess the systematic uncertainties in the significance values due to background subtraction. }
	\label{fig:histogram}       
\end{figure*}

We estimate the systematic uncertainties in our background estimation by measuring the significance distribution outside of the exclusion mask. 
Figure~\ref{fig:histogram} shows the distribution over the entire energy range, together with the best-fitting Gaussian function. 
The Gaussian has a width $(1.0830 \pm 0.0054)\sigma$, higher than the expected value of exactly 1$\sigma$. Its mean is $-(0.0421 \pm 0.0054)\sigma$, which is lower than the expected value of $0\sigma$. 
After correcting for this effect, emission is detected at significances of 7.2$\sigma$ and 6.3$\sigma$ for the eastern and western outer jets, respectively. 

\begin{figure*}
	\centering
	\includegraphics[width=0.9\textwidth]{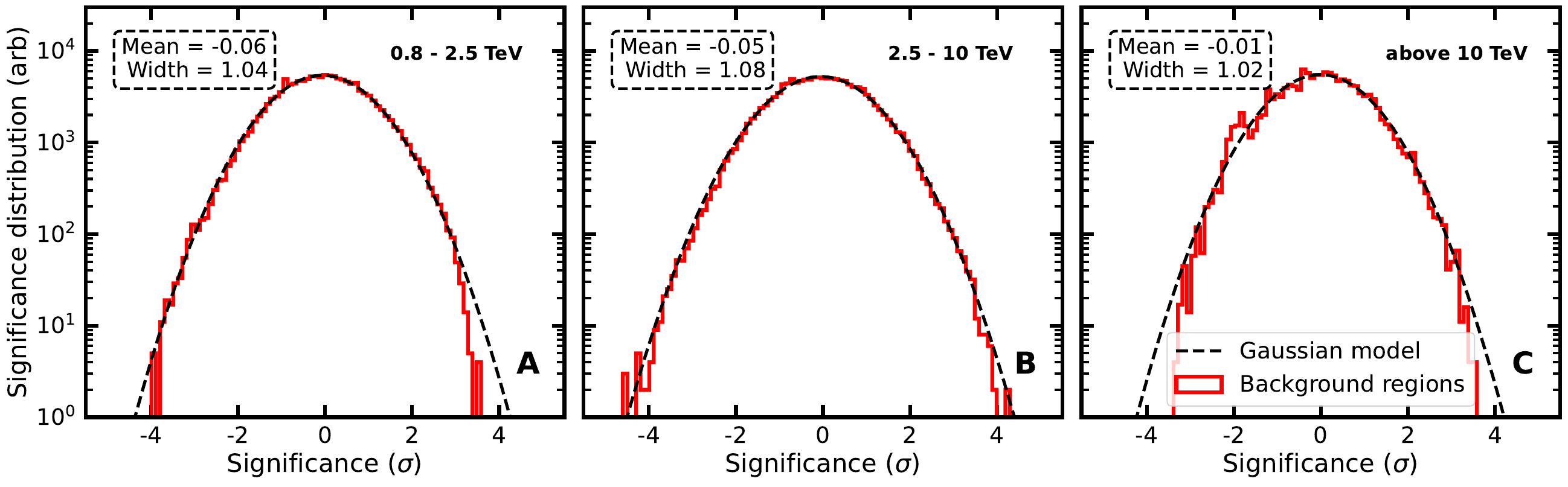}
	\caption{\textbf{Background significance distribution in different energy bands.} Same as Figure~\ref{fig:histogram} but split into the energy bands (\textbf{A}) 0.8 to 2.5~TeV, (\textbf{B}) 2.5 to 10~TeV and (\textbf{C}) above 10~TeV.}
	\label{fig:histogram_energy}       
\end{figure*}

For the significance maps in different energy (Figure~\ref{fig:significance-energy}), we derived equivalent distributions for each of the energy bands separately. The resulting histograms are shown in Figure~\ref{fig:histogram_energy}, with widths of $(1.0379\pm 0.0068)\sigma$, $(1.0718 \pm 0.0062)\sigma$ and $(1.018 \pm0.015)\sigma$ from low to high energy. The corrected maximum significance values when accounting for this effect are 4.2$\sigma$, 7$\sigma$ and 5.8$\sigma$ for the eastern jet from low to high energy and 4.5$\sigma$, 5.2$\sigma$ and 6.5$\sigma$ for the western jet in the same order.

\subsection*{Spectra of the jets}
Due to the energy-dependent morphology, assuming a single spatial model for the jets across all energies would introduce inaccuracies in the associated spectral model when fitting both components together. 
Therefore, we measured the spectra of the jets (after subtracting \jnoe) without assuming a spatial model by extracting the spectral information inside two elliptical regions (Figure~\ref{fig:large_significance}) large enough to completely contain the gamma-ray excess in each jet. 
Spectral models were fitted to the measured excesses in these regions using a maximum-likelihood framework. 
Models were compared via their test-statistic (TS) value, with the significance of the improvement in the description due to one additional parameter computed as $\sigma = \sqrt{\Delta \mathrm{TS}}$. 
For both jets the differential photon spectrum $\frac{dN}{dE}$ is described by a power-law $\phi_0 \left (\frac{E}{E_0} \right)^{-\Gamma}$ where $\Gamma$ the photon spectral index, $E_0$ the fixed reference energy and $\phi_0$ is the amplitude at the reference energy.
We find no significant ($>3\sigma$) evidence for curvature (1.5$\sigma$ and 0.7$\sigma$ for east and west) or an exponential cutoff (1.3$\sigma$ and 0.15$\sigma$) in the spectrum. 
The best-fitting parameters are listed in Table~\ref{tab:parameters-spectral}; both jets have consistent parameters within the uncertainties. 
Flux points were derived by fitting a normalisation parameter in each energy bin, assuming the best-fitting spectral shape derived from the wider energy range. The resulting flux and best-fitting model spectra are shown in Figure~\ref{fig:significance}.

\begin{table}[t!]
	\centering
	\caption{\textbf{Parameters of the power-law model fitted to the gamma-ray spectra of the jets.} Listed are the best-fitting and fixed parameters in the model: $E_0$ is the reference energy,  $\phi_0$ is the amplitude at the reference energy and $\Gamma$ is the photon spectral index.}
	\begin{tabular}{c | c c c}
		& $\phi_0$ & $E_0$  & $\Gamma$ \\
		&($10^{-13}$ TeV$^{-1}$ cm$^{-2}$ s$^{-1}$) & (TeV)\\
		\hline
		east  &$2.30 \pm 0.58_{\mathrm{stat.}} \pm 0.32_{\mathrm{syst.}}$  & 1 &  $2.19 \pm 0.12_{\mathrm{stat.}} \pm 0.12_{\mathrm{syst.}}$ \\
		\hline
		west & $2.83 \pm 0.70_{\mathrm{stat.}} \pm 0.39_{\mathrm{syst.}}$  &  1 & $2.40 \pm 0.15_{\mathrm{stat.}} \pm 0.13_{\mathrm{syst.}}$ \\
	\end{tabular}
	
	\label{tab:parameters-spectral}
\end{table}

\subsection*{Spatial properties of the energy-integrated gamma-ray emission}
\label{subsec:spatial}
The energy-integrated spatial properties of the gamma-ray emission were modelled using two elliptical Gaussian components. The elliptical description is preferred to a symmetrical Gaussian description by 5.8$\sigma$ and 3.5$\sigma$ for east and west, which in turn is preferred to a point-like description by 7.8$\sigma$ and 4.7$\sigma$ for east and west. 
The angle of the asymmetrical Gaussian model $\theta$ (degrees east from north) is fixed to that of the \ssftt \xray jets ($\theta=-19\degree$). 
Allowing the angle to vary during fitting of the model to the data results in $\theta=-16.2 \pm 3.5 \degree$ for the eastern excess, preferred by 0.74$\sigma$, and $\theta=-7.1 \pm 5.4 \degree$ for the western excess, preferred by 1.9$\sigma$. 
As neither is significant ($>3\sigma$), we keep the angle fixed to the value from \xray observations. 
The parameters of the best-fitting elliptical Gaussian model for each jet are presented in Table~\ref{tab:parameters-spatial}. The spatial extension is significant ($>3\sigma$) in both the major and minor axis directions of the ellipse.

\begin{table}[t!]
	\centering
	\caption{\textbf{Parameters of the elliptical model fitted to the spatial morphology.} Listed are the best-fitting and fixed parameters in the model: $l$ and $b$ are the Galactic longitude and latitude respectively of the central position, $r_{\mathrm{maj}}$ and $r_{\mathrm{min}}$ 
		are the major and minor axis standard deviation sizes and $\theta$ is the angle of the jets, which was fixed in the fit. The physical sizes for $r_{\mathrm{maj}}$ and $r_{\mathrm{min}}$ are calculated for a distance of 5.5~kpc~\cite{Blundell2004}. Uncertainties are both systematic and statistical.}  
	\resizebox{\textwidth}{!}{
		\begin{tabular}{c | c c c c c c}
			& unit  & $l$ & $b$  & $r_{\mathrm{maj}}$ & $r_{\mathrm{min}}$   & $\theta$  \\
			\hline
			east & deg &$39.875 \pm 0.018 $  & $-2.687 \pm 0.027 $ &  $0.205 \pm 0.035$ &  $0.044   \pm 0.014$ & -19 \\
			&   pc & &  &  $19.68 \pm 3.36$  &  $4.22 \pm 1.35$ & \\
			
			\hline
			west & deg & $39.564 \pm 0.013 $  &  $-1.853 \pm 0.027 $ & $0.134 \pm 0.036$ &  $0.046   \pm 0.015$ & -19  \\
			& pc  & &  &  $12.86 \pm 3.46$ &  $5.37 \pm 1.44$  & \\
			
	\end{tabular}}
	\label{tab:parameters-spatial}
\end{table}

\subsection*{Morphology in different energy bins}
\begin{figure}[]
	\centering
	\includegraphics[width=0.8\textwidth]{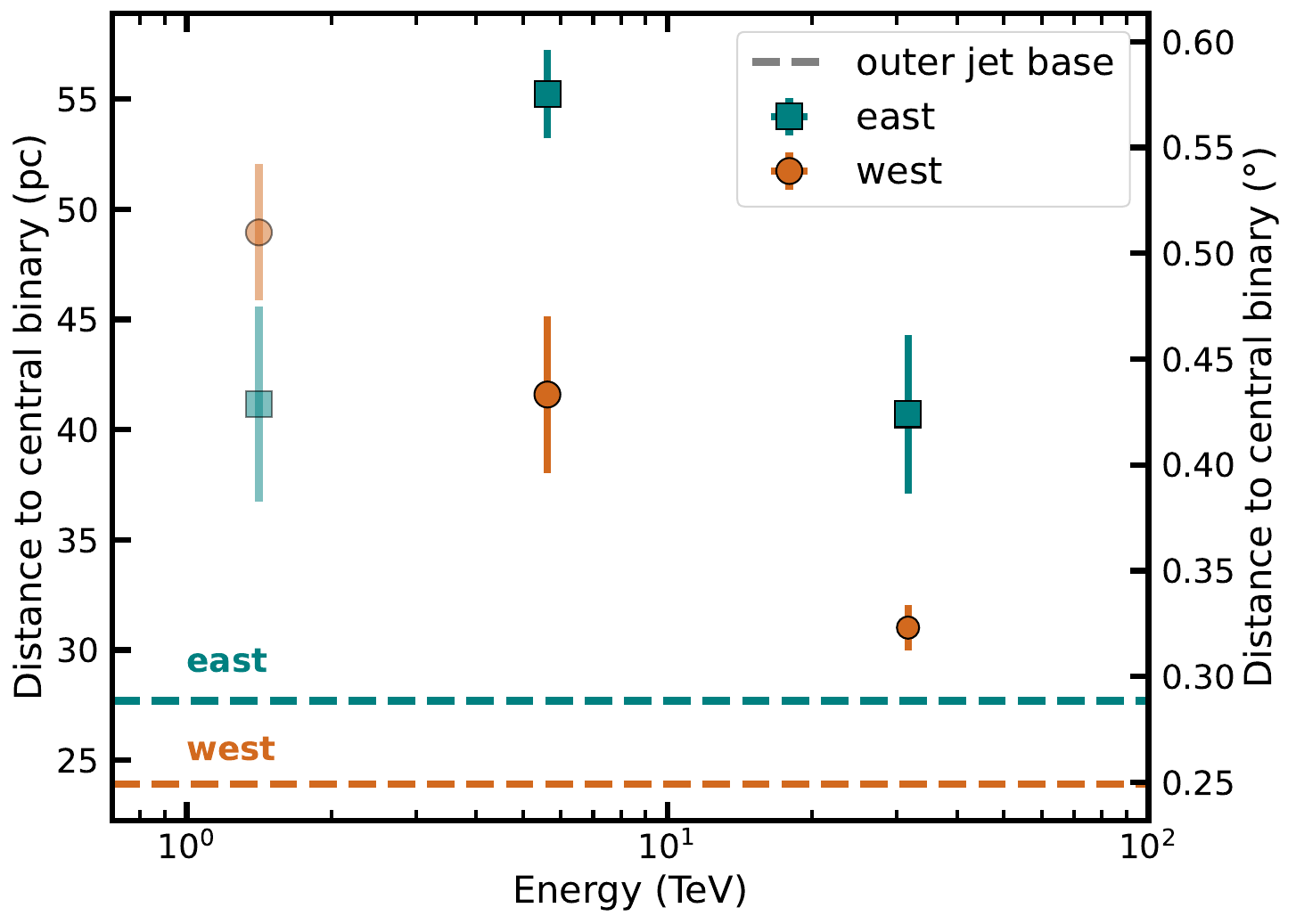}
	\caption{\textbf{Distance of the peak gamma-ray emission from the central binary. }
		The distance from the best-fitting position of the \gammaray emission centroid to the central binary is shown for both the eastern (green squares) and the western (orange circles) jets. In the energy range between 0.8 and 2.5~TeV, each jet is detected with a significance only above 4$\sigma$, which is why the measurements in this range are represented with a transparent point. Error bars indncate the combined statistical (1$\sigma$) and systematic uncertainties. The location of the base of the outer jets as inferred from \xray maps~\cite{Brinkmann1996} is marked with a dashed line for each side. The distances in parsec are calculated adopting a distance of 5.5~kpc to the system.~\cite{Blundell2004}}
	\label{fig:distances}
\end{figure}

Gamma-ray excesses are detected along both jets in all energy bins (Figure~\ref{fig:significance-energy}) with peak significance values (from low to high energy) of 4.4$\sigma$, 7.6$\sigma$, 5.9$\sigma$ for the eastern jet, and 4.7$\sigma$, 5.6$\sigma$ and 6.6$\sigma$ for the western jet. 
The morphology of the measured gamma-ray excesses was fitted separately in each energy band, to compare the resulting best-fit positions. 
The gamma-ray excesses are found to be significantly ($>3\sigma$) extended (not point sources) and are modelled with a symmetric Gaussian in most cases; however the western jet at $>$10~TeV is not significantly extended, so we used a point source model. 
The spectral parameters are fixed to those from the fitting to the full energy range (Table~\ref{tab:parameters-spectral}). 
We tested making these free parameters, which resulted in consistent spatial parameters in all cases [both with each other, and with the values from the full energy range (Table~\ref{tab:parameters-spectral})]. 
The distance from the position of the \ssftt binary at Galactic coordinates $l$=39.694\degree\xspace, $b$=--2.245\degree\xspace for each energy range is shown in Figure~\ref{fig:distances}. 
As a further check, the data were split into 3 bands with different energy boundaries and into different numbers (5, 10) of energy bands. 
In all cases, the resulting trend was consistent within the statistical uncertainties.

\begin{table}[t!]
	\centering
	\caption{\textbf{Results of spatial model fitting different energy ranges.} Same as Table~\ref{tab:parameters-spatial} but for three different energy ranges, each fitted separately. The distance from the model center to the nominal position of \ssftt $d_{\mathrm{\ssftt}}$ is listed both in degrees and in pc, the latter adopting a distance estimate of 5.5~kpc~\cite{Blundell2004}. Uncertainties are both systematic and statistical.}    \label{tab:energy-dep}
	\resizebox{\textwidth}{!}{
		
		\begin{tabular}{c | c c c c c c}
			side & energy  & $l$ & $b$ &$r$ & $d_{\mathrm{\ssftt}}$ & $d_{\mathrm{\ssftt}}$ \\
			& (TeV) &(deg) & (deg) & (deg) & (deg) & (pc)\\
			\hline
			east  & $0.8\text{ to }2.5$   & $39.913 \pm 0.044 $ &  $-2.614 \pm 0.047$ & $0.125\pm0.022$ &$0.428 \pm 0.046$  &  $41.148 \pm 4.424$ \\
			& $2.5\text{ to }10$   & $39.924 \pm 0.018 $ &  $-2.772 \pm 0.021$ &$0.085 \pm 0.015$&$0.575   \pm 0.021$ &  $55.212\pm 2.007$ \\
			&  above $10$   & $ 39.840 \pm 0.031 $ &  $-2.643 \pm 0.038$ & $0.013 \pm 0.029$ &$0.424  \pm 0.037$   &  $40.693 \pm 3.593$ \\
			\hline
			west  & $0.8 \text{ to } 2.5$   & $39.537 \pm 0.024 $ &  $-1.759 \pm 0.033$ &$0.080\pm 0.016$ & $0.510   \pm 0.032$   &  $48.946 \pm 3.089$ \\
			& $2.5 \text{ to } 10$   & $39.582 \pm 0.024 $ &  $-1.826 \pm 0.037$ & $0.095 \pm 0.018 $ &$0.433   \pm 0.037$&  $41.590 \pm 3.552$ \\
			&  above $10$  & $39.560 \pm 0.010 $ &  $-1.951 \pm 0.011$ & - &$0.323   \pm 0.011$  &  $31.015 \pm 1.038$
			
	\end{tabular}}
	
\end{table}

\subsection*{Gamma-ray spatial flux profiles}
Spatial flux profiles along the jets in different energy bands were derived from the \hess data. 
We used them to visualise the contamination from \jnoe (Figure~\ref{fig:profiles-jnoe}) and for comparison with one-dimensional theoretical models (Figure~\ref{fig:profile}). 
We constructed the profiles by defining an axis passing through both jets and the central binary as the line between the Galactic coordinates ($l$, $b$)=(40.246\degree\xspace,-3.695\degree\xspace) and (39.340\degree\xspace,-1.295\degree\xspace) (Figure~\ref{fig:large_significance}). 
We defined a rectangular box of full width 0.7\degree\xspace along this axis and sliced it into 19 perpendicular boxes of height 0.14\degree\xspace. 
The height of the boxes is chosen as a compromise between having enough signal in each slice and sampling the spatial shape of the gamma-ray emission, for comparison the 68\% containment radius of \hess PSF is $\sim$0.07\degree\xspace. 
The definition of these boxes does not consider the positions of, or distances between the emission centroids in the different energy bands (Table~\ref{tab:energy-dep}). 
In each of these boxes the excess counts were computed and a normalization was fitted assuming a power-law spectral shape with index 2.3, the mean of the measured indices of the eastern and western jets. 
Using instead the best-fitting value for either of the jets, of 2.2 and 2.4 for east and west, respectively (Table~\ref{tab:parameters-spectral}), gives consistent results. The resulting integrated flux profiles are shown in Figures~\ref{fig:profile} and~\ref{fig:profiles-jnoe}. 
The outlier flux point seen at around 0.7\degree\xspace towards the eastern side above 10~TeV is the result of a background fluctuation outside the jet region (Figure~\ref{fig:significance-energy}C). Reducing the width of the box could prevent this outlier, but also excludes some of the gamma-ray flux, especially at the lower energies. 
Therefore, we choose to keep the box wide enough to contain the gamma-ray emitting regions entirely.

\subsection*{Search for periodic variability}
Fermi J1913+0515 is a GeV gamma-ray source reported to pulsate with a period consistent with the precession of the \ssftt jets~\cite{Li2020}. The position of Fermi~ J1913+0515 and its uncertainty are shown in Figure~\ref{fig:significance}, with no significant emission at that location. However, due to the periodic nature of Fermi J1913+0515, it is possible that any TeV emission would only be detectable around certain phase ranges, like in the GeV range. We investigated the presence of TeV emission from Fermi J1913+0515 by phase-folding the \hess observations. We adopted the jet precession period to be 162.250 days~\cite{Davydov2008} and a starting time $T_0$ of JD 2443508.4098~\cite{Li2020} and derived a corresponding phase for each \hess observation. We separated the observations into 8 phase bins and repeated the data reduction process for each group, resulting in a full map of the entire region in each case. 
No significant emission was found in the vicinity of Fermi~J1913+0515 in any of the phase ranges considered, both for the full energy range or the three energy ranges used in Figure~\ref{fig:significance-energy}. 
No significant emission was found either when considering 4 and 2 bins of phase. 
A non-detection in the TeV range is consistent with the observation of energies only up to 10~GeV from Fermi~J1913+0515~\cite{Li2020}. 
No significant phase trend was found for any other part of the \ssftt system in any of the phase or energy ranges considered.

While a sub-threshold excess in the region between Fermi~J1913+0515 and the \xray contours can be seen in the phase-integrated significance map shown in Figure~\ref{fig:significance}, no phase trend or variation is observed at this location. When split into different bins of phase, the sub-threshold excess at this location disappears, indicating no apparent relation with the jet precession phase. Additionally, when the jet emission is modelled with an elliptical Gaussian, no significant (or near-threshold) emission is left at that location, meaning that our measured TeV gamma-ray counts from the eastern jet region are compatible with a single component aligned with the jet axis.

\subsection*{Upper limits from the central source and e3 regions}
\begin{figure}[]
	\centering
	\includegraphics[width=0.8\textwidth]{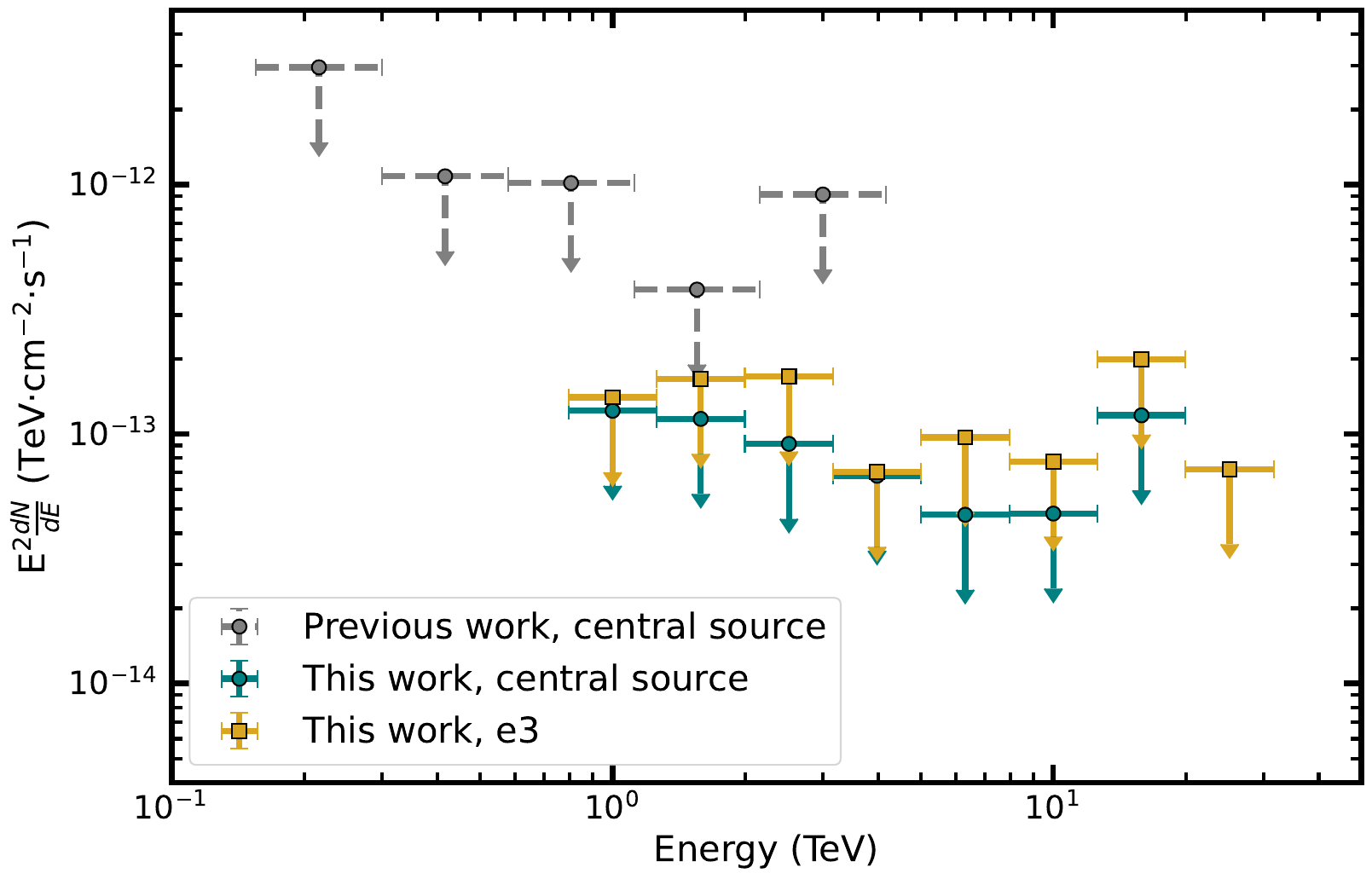}
	\caption{\textbf{Upper limits for undetected regions. } 95\% confidence level flux upper limits (downward arrows) derived at the position of the central binary (teal circles) and the e3 region (yellow squares). Gray circles are limits on the central source from previous observations~\cite{HESSMAGIC2018}.}
	\label{fig:central_ul}
\end{figure}

No significant ($>5\sigma$) excess of TeV gamma rays is detected at either the position of the binary system or at the  \xray jet termination region e3 (Figure~\ref{fig:significance}A). 
Differential upper limits for both regions were computed assuming a spectral index of 2.7 following previous methods~\cite{HESSMAGIC2018}. 
The radius of the circular region where the points are extracted is 0.07\degree\xspace and 0.2\degree\xspace for the central and e3 regions, respectively, motivated by the size of the \xray emission in each region. 
The resulting limits are shown in Figure~\ref{fig:central_ul}, together with the previously most constraining upper limits~\cite{HESSMAGIC2018} for the central source. The non-detection of TeV emission confirms that the \xray emission from these two regions is predominantly of thermal origin~\cite{Brinkmann1996,Marshall2002}.

\subsection*{Systematic uncertainties in model parameters}
The systematic uncertainties in the model parameters are calculated with a Monte Carlo-based approach, in which the IRFs are randomly varied to generate random pseudo-datasets based on the best-fitting spatial and spectral models described above.
These datasets are then re-fitted using the original, unmodified IRFs. 
The obtained spread in the fitted parameters then reflects their combined statistical and systematic uncertainty. 
This procedure is extensively described elsewhere~\cite{Collaboration2023}. 
The resulting systematic uncertanties are consistent with previous estimates of the \hess systematic errors~\cite{Aharonian2006}. The contamination of \jnoe into the jet regions was assessed by not including that model component in half of the generated pseudo-datasets. 
The same procedure was used when calculating systematic errors for flux points. In this case, the resulting systematic uncertainty is of the same magnitude as the statistical one at energies around 1~TeV, and quickly becomes negligible at higher energies.
For certain fitted parameters, such as the eccentricity and width of the elliptical Gaussian components, the distribution is not broadened by the systematic effects considered in this estimate. 
This is also the case for the best-fitting Galactic coordinates of the source components. 
However, these parameters may be affected by other systematic effects neglected in the Monte Carlo approach. 
In particular, the source positions are subject to a systematic uncertainty of the pointing position of the H.E.S.S. telescopes, which is of the order of 10$^{\prime\prime}$~-~20$^{\prime\prime}$~\cite{GillessenThesis}. 
This value is around a factor 10 lower than the statistical uncertainty on the positional measurements (see Table~\ref{tab:parameters-spatial}). 
We accounted for this source of systematic uncertainty with a 20$^{\prime\prime}$ systematic uncertainty included in the best-fitting coordinates presented in Tables~\ref{tab:parameters-spatial} and~\ref{tab:energy-dep}.

\subsection*{Multi-wavelength data}
We used multi-wavelength (MWL) observations of the outer jets of \ssftt to determine the spectral energy distribution (SED) of both jets. 

\subsubsection*{Radio}
The outer jets have not been detected in the radio band (hundreds of km to mm wavelengths), a range in which instead the shell of  W\,50 is a bright source. 
We used data from the 11~cm Effelsberg radio telescope survey~\cite{Reich1984,Reich1990, Furst1990} are used to determine the total flux inside the region where the gamma-ray spectra are extracted. 
This provides an upper limit of $7.084 \cdot 10^{-14}$ TeV s$^{-1}$ cm$^{-2}$ and $1.367 \cdot 10^{-13}$ TeV s$^{-1}$ cm$^{-2}$ to the radio flux coming from the eastern and western jets respectively, as they must at least be fainter than W\,50.

\subsubsection*{\xray}
\label{subsubsec:xray}
\begin{figure}[]
	\centering
	\includegraphics[width=0.8\textwidth]{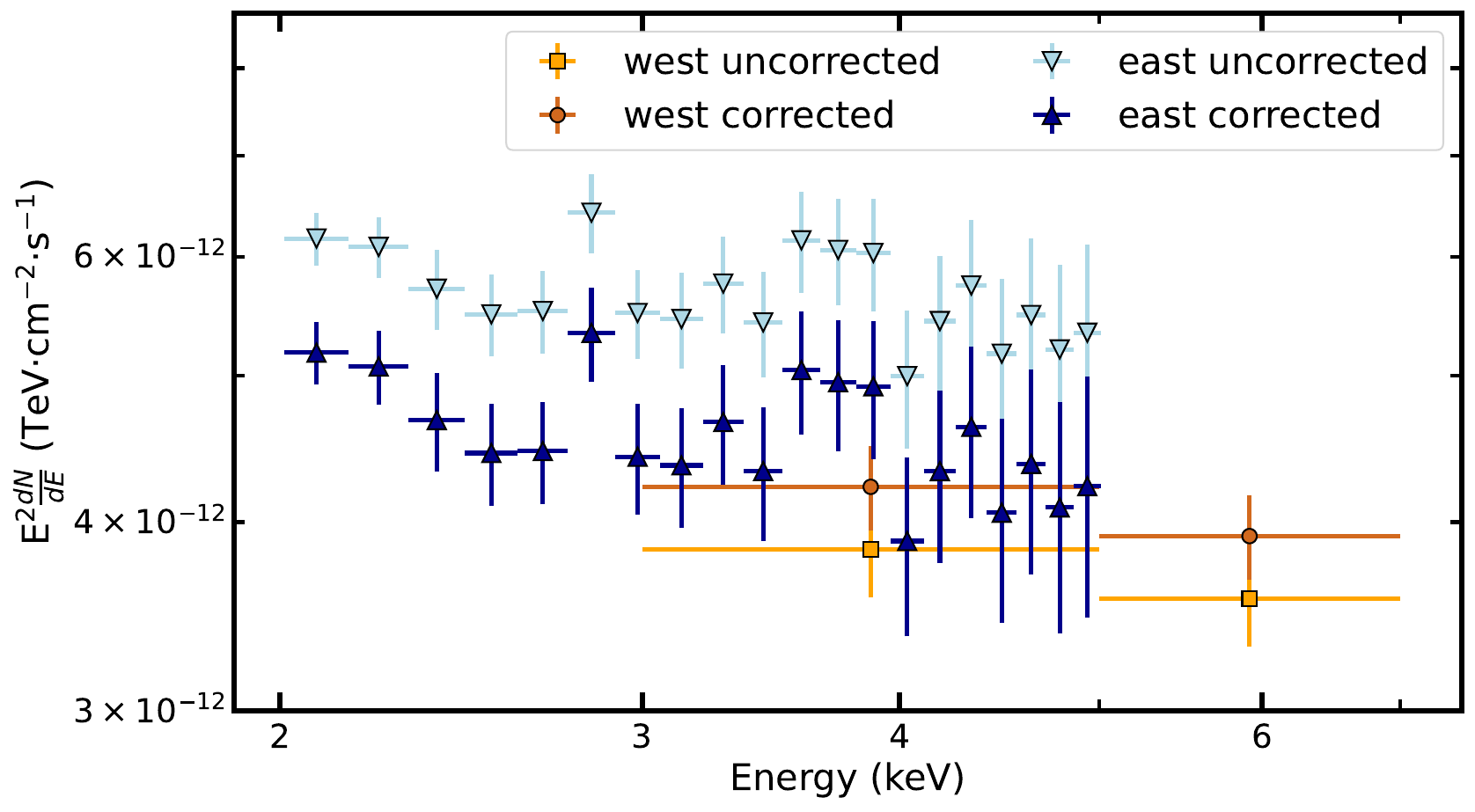}
	\caption{\textbf{\xray flux points. }The \xray flux of the eastern jet derived from XMM-Newton data is shown before (light blue upside down triangles) and after (dark blue triangles) the corrections described in the text. Same for the Chandra data of the western jet but with orange circles and yellow squares respectively. Error bar indicate statistical (1$\sigma$) uncertainties.}
	\label{fig:xray}
\end{figure}

\xray flux was determined from data taken by the X-ray Multi-Mirrors-Newton (XMM-Newton)~\cite{Brinkmann2007} and Chandra~\cite{Kayama2022} space telescopes for the eastern and western jet, respectively. 
The flux points were extracted from the same spatial regions used for the TeV measurements, with the limitation of the different field of view from the different instruments. 
Only data above 2~keV is used to minimise the contribution of thermal \xray emission and the effect of absorption. 
In the case of the Chandra data the effect of the small field-of-view is taken into account with a correction factor of order 10\%, derived from the surface brightness of a soft \xray image of the jets~\cite{Brinkmann2007}. 
The \xray emission from the eastern jet includes that of the bright region labelled e2 (Figure~\ref{fig:significance}A). 
The relative brightness of this region with respect to the rest of the \xray emission from the eastern jet suggests that it might be the result a local enhancement of the magnetic field. 
Because we are interested in the average magnetic field of the jets, we subtract the spectrum of this region~\cite{Safi-Harb2022}; we discuss the effects of this correction below. The resulting \xray flux points for both jets before and after the corrections are shown in Figure~\ref{fig:xray}.

\subsubsection*{GeV gamma rays}
The presence of a bright pulsar in the field of view and uncertainties on the Galactic diffuse emission make the \ssftt region difficult to study at GeV energies, with different studies
reaching conflicting conclusions~\cite{Bordas2015, Sun2019, Xing2019, Rasul2019}. 
Evidence for ($\sim 4\sigma$) an emission excess in the \textit{Fermi} Large Area Telescope (\textit{Fermi}-LAT) data has been reported~\cite{Fang2020} near the outer western jet, but not spatially coincident with it. The 95\% containment upper limits for both the western and eastern jets~\cite{Fang2020} are reproduced in Figures~\ref{fig:SED} and~\ref{fig:age}.

\subsection*{Multi-wavelength spectral model}
We modelled the MWL spectral energy distribution of each jet using a single electron population injected continuously at the base of the outer jets. 
This model is fitted to the MWL data described above and the \hess data assuming two one-zone (eastern and western) scenarios. 
Transport effects within the jets are not included, because we assume particles cool at the same rate everywhere in the jet. 
We use the gamera package~\cite{Hahn2015,Hahn2022} to model the temporal evolution of the electron distribution and the resulting radiation. 
The injected electron spectrum is parameterised as a power-law with index $\Gamma_e$ and an exponential cutoff at $E_{\mathrm{cut}}$. 
Synchrotron radiation produced by electrons with a spectral index $\Gamma_e$ is expected to follow a photon spectral index of ($\Gamma_e$+1)/2. 
The measurement of an \xray photon index of roughly 1.5 at the base of both the western~\cite{Kayama2022} and eastern~\cite{Safi-Harb2022} outer jets consequently implies $\Gamma_e = 2$. Therefore, we fix the value of the electron spectral index to $\Gamma_e = 2$ in our fit.
The power injected in electrons above 10~MeV is parameterised as a fraction $\alpha$ of the jet kinetic power, the latter taken to be 10$^{39}$erg~s$^{-1}$. 
The injected electron spectrum is assumed to have the same parameters for both jets. 

\begin{figure}[]
	\centering
	\includegraphics[width=0.97\textwidth]{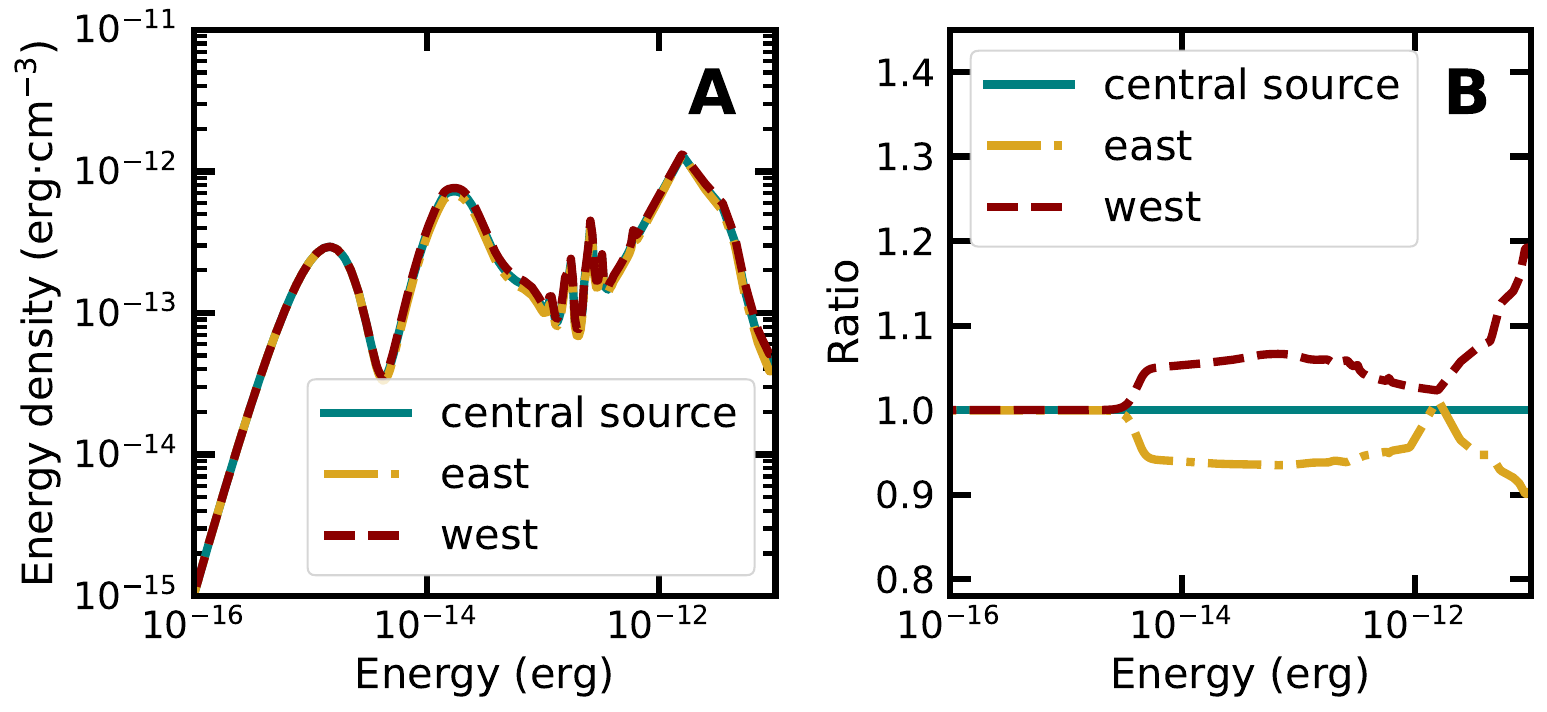}
	\caption{ \textbf{Model of the ambient radiation fields. }\textbf{A:} Model energy density of the background radiation fields~\cite{Popescu2017} at the position of the central binary (teal solid line), the eastern (yellow dash-dotted line) and the western (red dashed line) jets. \textbf{B: }Energy densities shown in panel \textbf{A} normalised by the model at the central binary highlighting the differences between the two jets.}
	\label{fig:radiation}
\end{figure}

\begin{figure}[]
	\centering
	\includegraphics[width=0.97\textwidth]{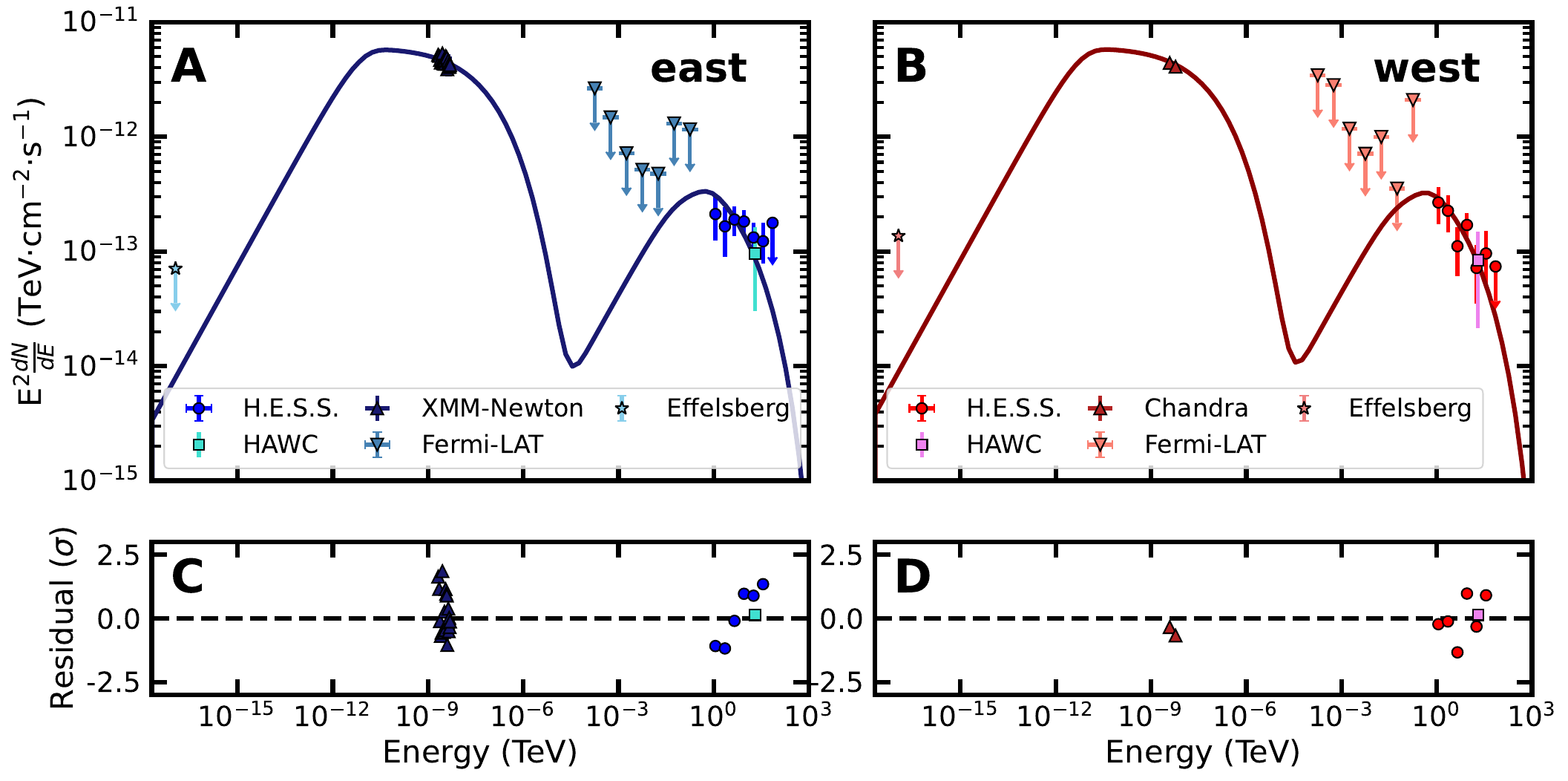}
	\caption{\textbf{Multi-wavelength spectral energy distribution of the \ssftt outer jets and best-fitting model. } Flux points and upper limits (downward arrows) derived from the gamma-ray emitting regions in the (\textbf{A}) eastern and (\textbf{B}) western outer jets in the radio (stars), \xray (triangles), GeV gamma-rays (upside-down triangles) and TeV gamma-rays including both this work (circles) and previous observations by the High-Altitude Water Cherenkov Observatory (HAWC, squares). The best-fitting model for each of the jets is shown with a solid line. The residual is shown in panels \textbf{C} and \textbf{D}.}
	\label{fig:SED}
\end{figure}

The injected electrons lose energy via two processes: synchrotron emission in a uniform magnetic field $B$ and IC scattering of photons from a uniform ambient radiation field. 
The photon emission from the binary peaks in the ultra-violet (UV) range~\cite{Fabrika2004}; IC scattering on UV photons is severely suppressed due to the Klein-Nishina effect~\cite{Blumenthal1970}, is included in the gamera package implementation of IC scattering. For TeV gamma-ray emission, far-infrared (FIR) photons provide instead the dominant scattering field. 
At distances of more than 25 pc from the binary, the dominant FIR photon field is not that of the binary or W\,50\cite{Band1987} but the diffuse IR background. 
We use the combination of an axisymmetric Galactic model~\cite{Popescu2017} derived from observations at similar distances from the Galactic Centre as the outer jets and the cosmic microwave background as target field. The energy densities of the assumed ambient fields are shown in Figure~\ref{fig:radiation}. 
The radiation fields differ by less than 10\% at the locations of each jet with respect to the one at the location of the central binary. The synchrotron photons produced by the electrons would in principle contribute to the target field (Synchrotron-self-Compton~\cite{Atoyan1996}), but this contribution is negligible as inferred from the predicted synchrotron flux (Figure~\ref{fig:SED}).

The models were fitted to the TeV and \xray data for each jet, using a maximum likelihood approach, and considering the GeV and radio upper limits. 
For simplicity, the magnetic field strength is the only parameter allowed to differ between the eastern and western sides. 
The best-fitting values for the parameters are shown in Table~\ref{tab:model-parameters}. 
We do not find significant ($>3\sigma$) evidence for an exponential cutoff in the electron spectrum, meaning that if such a feature is present, it must be at higher electron energies than those probed by \hess observations. 
Therefore we derive a 68\% C.L. lower limit for the cutoff energy in the electron spectrum of $E_{\mathrm{cut}}>$200~TeV.
The power required to reproduce the observed emission is around 0.13\% of the jet kinetic power of 10$^{39}$~erg~s$^{-1}$. 
The best-fitting magnetic field strenghts are roughly 19 and 21~\textmu G for the eastern and western jet, respectively, in agreement with estimates derived from \xray observations~\cite{Safi-Harb2022}. 
For reference, the Poynting flux of the jet assuming a velocity $\sim0.1c$ and these values of the magnetic field is $\sim 10^{37}$~erg~s$^{-1}$, around one order of magnitude more than the power in the relativistic electrons. 
The resulting MWL spectral model is shown in Figure~\ref{fig:SED} for both jets.\\

\begin{table}[h]
	\centering
	\caption{\textbf{Best-fitting parameters from the multi-wavelength emission model.} Best-fit values of the injected electron spectral index ($\Gamma_e$), exponential cutoff (E$_{\mathrm{cut}}$), power as a fraction of the jet kinetic power ($\alpha$) and the magnetic field ($B$). Uncertainties are statistical only. The third column indicates whether each parameters was shared for bot jets, or allowed to vary between them.}
	\begin{tabular}{c |  c c c c}
		& east & west  & shared & fixed \\
		\hline
		$\Gamma_e$ & 2 & 2 & yes & yes \\
		$E_{\mathrm{cut}}$ (TeV) & \multicolumn{2}{c}{$>$200} &  yes & no \\
		$\alpha$  & \multicolumn{2}{c}{($1.287\pm 0.029$)$\cdot 10^{-3}$}  & yes & no \\
		$B$ (\textmu G) & 19.5$\pm$2.7 & 21.1$\pm$1.8 & no & no \\
	\end{tabular}
	
	\label{tab:model-parameters}
\end{table}

If the \xray flux from the e2 region in the eastern jet is not subtracted (Figure~\ref{fig:xray}), the parameter $\alpha$ varies between the eastern and western jets, resulting in $\alpha_{\mathrm{east}} = (1.597\pm 0.027)\cdot 10^{-3}$, $B_{\mathrm{east}} = 20.0\pm2.6$~\textmu G, $\alpha_{\mathrm{west}} = (1.184\pm 0.073)\cdot 10^{-3}$ and $B_{\mathrm{west}} = 20.8\pm1.8$~\textmu G. 
While these values of the magnetic field are consistent (within the uncertainties) with those reported in Table~\ref{tab:model-parameters}, including the emission from the e2 region requires more power to be injected into the electrons on the eastern side. Providing more power to the eastern side would not affect our conclusions, so for the subsequent discussions we refer to the model derived from the corrected \xray data.

\begin{figure}[]
	\centering
	\includegraphics[width=0.9\textwidth]{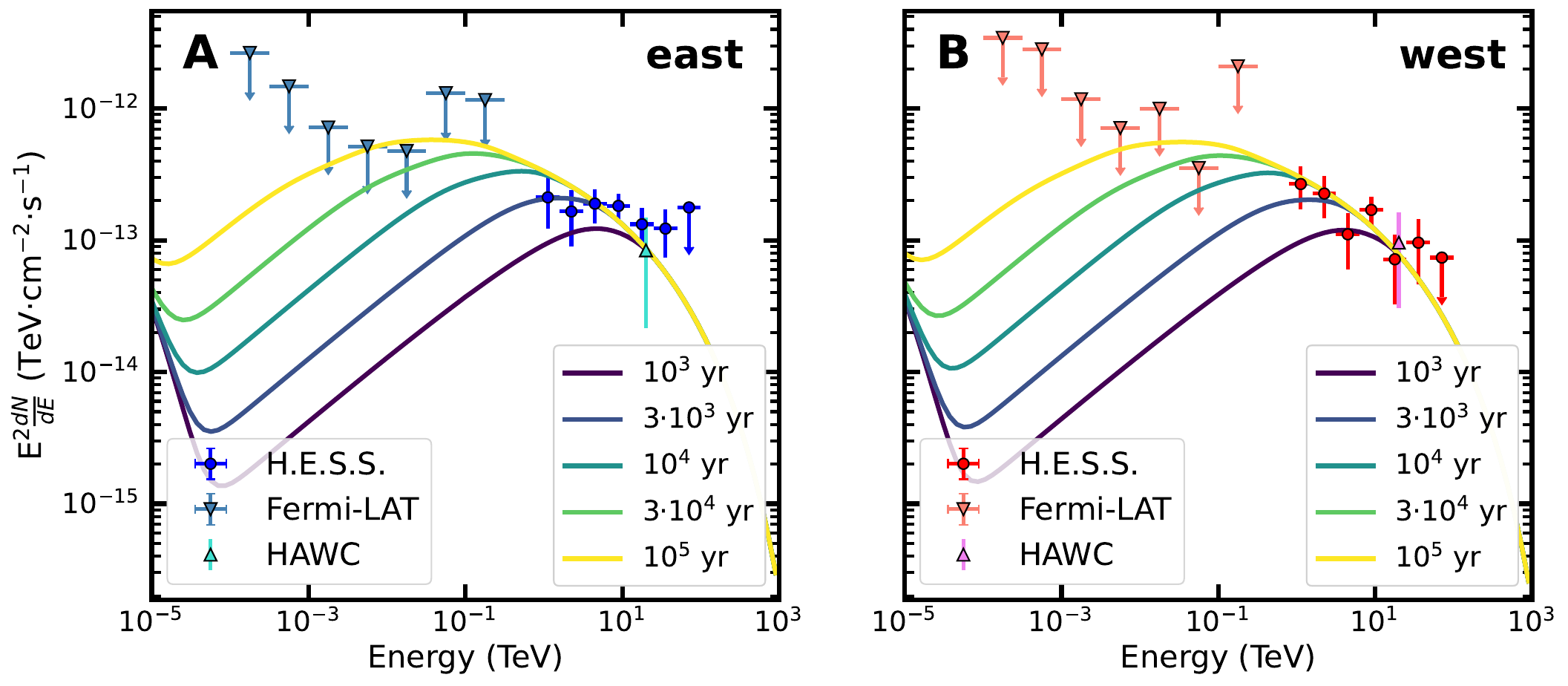}
	\caption{\textbf{Model predictions for different values of the injection time.} Coloured lines show emission as a function of the assumed injection time (colour bar) for the eastern (\textbf{A}) and western (\textbf{B}) jets, assuming the parameters listed in Table~\ref{tab:model-parameters}.
		TeV measurements and GeV upper limits are the same as in Figure~\ref{fig:SED}.}
	\label{fig:age}
\end{figure}

The combination of the injection index required by the \xray data, the GeV upper limits and the steepness of the \hess spectra constrains for how long the electrons are injected (Figure~\ref{fig:age}), which in turn constrain the age of the jets (or its recent activity).
Assuming continuous injection with the parameters listed in Table~\ref{tab:model-parameters}, injection is constrained to last more than 1\,000~yr, but less than 30\,000~yr, with an age of between 3\,000 and 10\,000 yr yielding better agreement with the observations (Figure~\ref{fig:age}). 
This result is consistent with previous theoretical and numerical studies, which placed the age of the W50/\ssftt complex, ranging between 10\,000-100\,000~yr, e.g.~\cite{Goodall2011a,BowlerKeppens2018}.

\subsection*{Comparison with \xray morphology}
\begin{figure}[]
	\centering
	\includegraphics[width=0.97\textwidth]{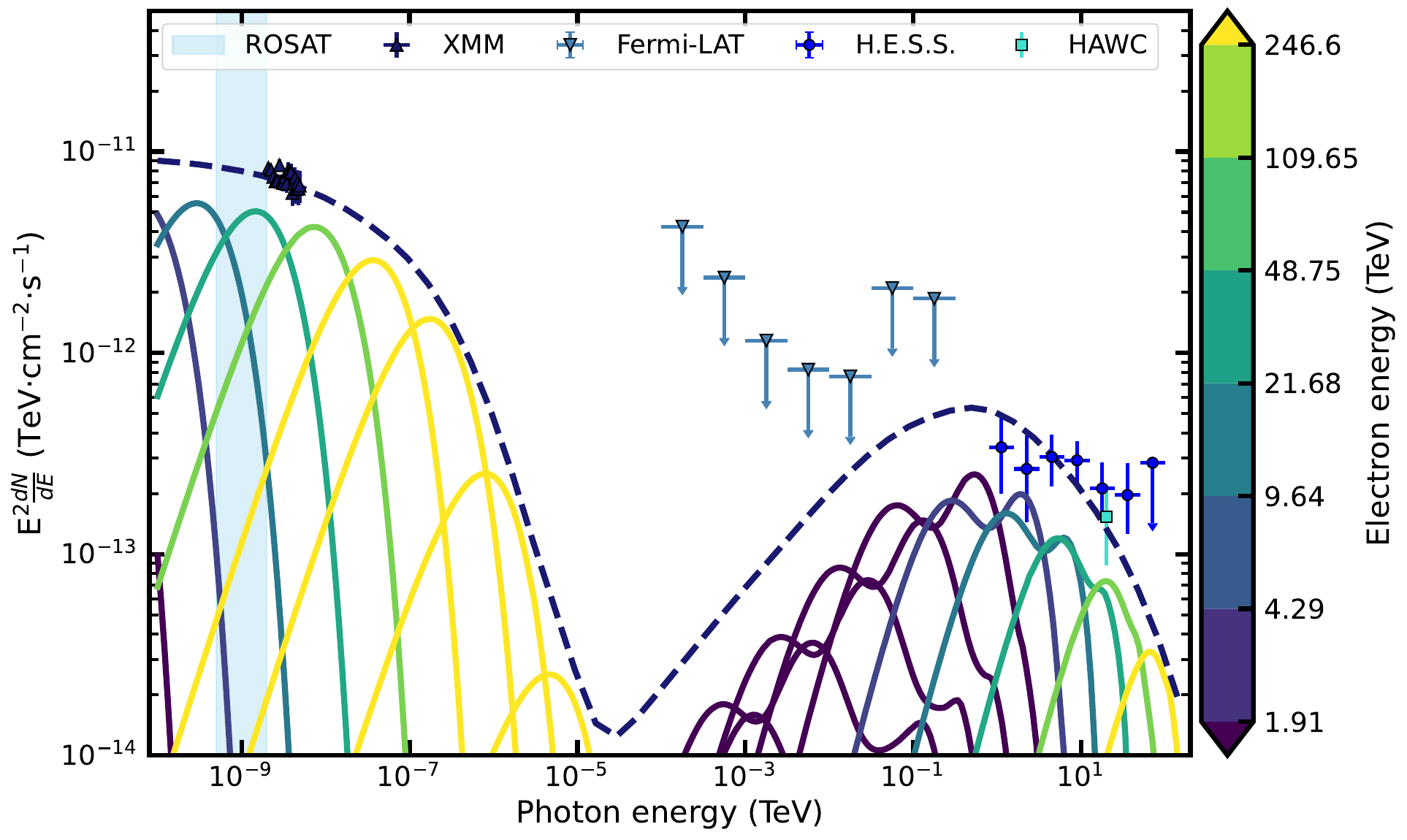}
	\caption{\textbf{Contribution from electrons of different energies to the eastern SED model.} Coloured solid lines show emission from electrons with different energies (colour bar). The range of electron energies depicted was chosen to overlap with the \hess observations. The model parameters used are listed in Table~\ref{tab:model-parameters}. The ROSAT energy range is indicated with a blue band. Flux points, upper limits and the total SED (dashed blue line) are the same as in Figure~\ref{fig:SED}A.}
	\label{fig:electron_energies}
\end{figure}

To compare the observed \gammaray morphology to the \xray morphology, we focus on the eastern jet using the \xray images covering the entire eastern jet in a wide range of photon energies from the Roentgensatellit (ROSAT)~\cite{Brinkmann1996} and XMM-Newton~\cite{Brinkmann2007, Safi-Harb2022} space telescopes. 
The morphology of the \xray emission from the eastern jet is also energy-dependent~\cite{Safi-Harb2022}, with predominantly hard (2 to 12~keV) \xray emission detected from the e1 region, a mixture of soft (0.5 to 2~keV) and hard emission from the e2 region and predominantly soft thermal \xray emission from the e3 region. 
Comparing the \gammaray morphology to the \xray morphology therefore requires a choice of which \xray energies to consider. 
We do so based on the model for the emission described above (Table~\ref{tab:model-parameters}). 
Figure~\ref{fig:electron_energies} shows the model contribution of electrons of different energies to the total SED. 
The electrons responsible for the \xray emission in the XMM-Newton energy range (2 to 7~keV) mostly produce gamma rays of energies above 10~TeV. 
The lower energy (0.5 to 2~keV) \xray emission detected by ROSAT is mostly produced by lower energy electrons, which are responsible for the remaining \gammaray emission measured by \hess below approximately 10~TeV. 
We therefore constructed two spatial templates using flux maps measured by ROSAT in the 0.5-2~keV band and XMM-newton in the 2-7~keV band. 

The spatial templates were used as a model that is fitted to the morphology of the \gammaray excess in the relevant energy range. 
This was done by multiplying the templates by a flux normalisation factor, which is a free parameter. Besides this template, an additional model component for \jnoe was used, also with the flux normalisation as a free parameter. 
The ROSAT template was fitted to the \hess data below 10~TeV and the XMM-Newton template was fitted to the \hess data above 10~TeV. 
In both cases, the residual map contained no excesses with significance greater than 5$\sigma$.
However, there is evidence for ($\sim 4\sigma$) an excess in the residual map in the lower energy model around the e1 position, likely due to the e2 region being several times brighter than the surrounding emission in the ROSAT image. 
Less significant emission can be seen in the residual of the higher energy model, where sub-threshold ($\sim 2.5\sigma$) positive and negative excesses are found around e1 and e2 respectively. 
If the ROSAT template is fitted to the entire \hess energy range, a significant ($>5\sigma$) excess remains at the base of the outer jet around e1, because the emission from that region arises predominantly from the highest energy electrons. 
We conclude that the observed energy-dependent morphology in the \xray and \gammaray bands are consistent with a shared origin due to the acceleration of particles at the base of the outer jet and their transport in the jet flow. 
The observed sub-threshold discrepancies may be explained with a modest local increase of the magnetic field in the e2 region.

\subsection*{Shock acceleration timescale}

\begin{figure}[]
	\centering
	\includegraphics[width=0.5\textwidth]{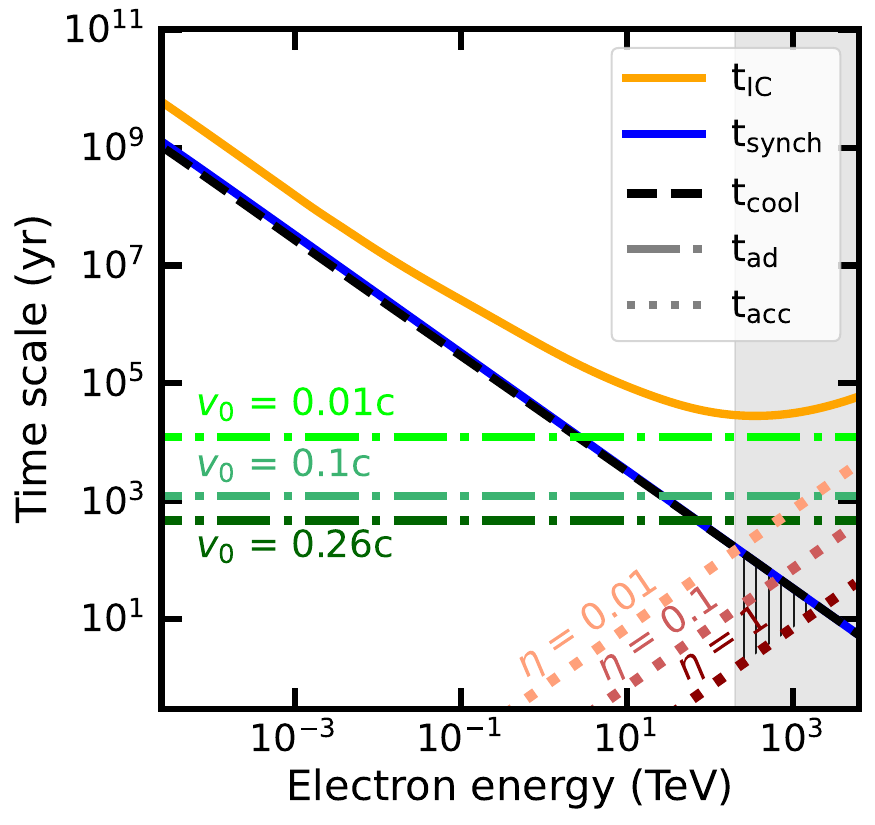}
	\caption{\textbf{Calculated cooling and acceleration timescales.} The cooling timescale t$_\mathrm{cool}$ is plotted as a function of electron energy (black dashed line) assuming the best-fitting values for the magnetic field (Table~\ref{tab:model-parameters}) and the same ambient radiation field as used in the MWL modelling. The contributions from IC (t$_\mathrm{IC}$) and synchrotron (t$_\mathrm{synch}$) losses are shown in orange and blue, respectively. Plotted are the values for the eastern jet; the western jet is almost identical. Dotted lines display the acceleration timescales (t$_\mathrm{acc}$) for different values of the efficiency $\eta$, assuming a jet velocity prior to the shock of $v_1 = 0.26c$. The maximum electron energy range allowed by the lower limit on the cutoff energy of the electron spectrum is indicated with a grey band. The region defined by the combination of this energy range, the cooling time and acceleration time is indicated with vertical hatching.
		Adiabatic loss time scales (t$_\mathrm{ad}$) assuming a jet with constant velocity, are depicted with green dot-dashed lines for different values of $v_0$, a proxy for the advection time. For a decelerating constant-density jet or a jet moving with both constant section and velocity, adiabatic losses are negligible. }
	\label{fig:cooling}
\end{figure}

We consider the timescale requirements for electrons to be accelerated at a shock located at the base of the outer jet.
The acceleration timescale as a function of energy $t_\mathrm{acc}$ is plotted in Figure~\ref{fig:cooling}, computed assuming test-particle diffusive shock acceleration theory~\cite{Drury1983}:
\begin{equation}
	\label{Eq:tacc}
	t_\mathrm{acc} = \frac{3}{u_1-u_2}\left(
	\frac{D_1}{u_1} + \frac{D_2}{u_2}
	\right) \approx \frac{8}{\eta}\frac{D_\mathrm{Bohm}}{u_\mathrm{1}^2},
\end{equation}
where $u_1$, $u_2$, $D_1$ and $D_2$ are the velocities in the shock frame and diffusion coefficients in the upstream and downstream media, respectively. The parameter $\eta$ is the ratio of an electron's mean free path to its gyroradius ($r_g$), a measure of the scattering efficiency with respect to idealised magnetised transport where $D=\frac{1}{\eta}D_\mathrm{Bohm}=\frac{1}{3\eta}r_gc$. Equation~\ref{Eq:tacc} assumes approximately equal upstream and downstream residence times. 
The case with $\eta = 1$ corresponds to the limiting minimum diffusion coefficient for magnetised transport (i.e. for which particles can still be considered to undergo helical trajectories) and therefore sets a lower limit on the acceleration time~\cite{Lagage1983}. 
Our emission model (Table~\ref{tab:model-parameters}) set a lower limit for the energy cutoff of the electron spectrum of $>$200~TeV. 
This implies values of $\eta\,(u_1/0.26 c)^2 \gg 0.01$ for electrons to compete with radiative cooling losses. 
Upstream velocities $u_1$ much less than 0.1$c$ would require $\eta>1$, implying non-magnetised transport. In this regime, additional care is needed in applying Equation~\ref{Eq:tacc}, but in general would make the inferred energies unreachable.

\subsection*{The effect of particle transport}
We interpret the observed energy-dependent position of the gamma-ray emission in the jets of SS 433 as a consequence of the combination of particle cooling timescales (Figure~\ref{fig:cooling}) and advection with the jet flow, and thus can be used to constrain the internal dynamics of the outer jets. 
We test this assumption using a one-dimensional Monte Carlo simulation which models the transport of particles via the combined action of advection and diffusion. 
We used gamera to calculate the radiation and cooling of particles as they are transported. 
The simulation was run from some initial time $t = 0$ to an assumed duration of the electron injection $t = t_{\mathrm{inj}}$. 

We define the coordinate $z$ as the position along the jet axis. At each time-step $t_{\mathrm{step}}$, electrons are injected at the base of the outer jets, assumed to be $z = 0$ with a spectral shape described by a power-law with an exponential cutoff. 
The parameters of the injected electron distribution and the value of the magnetic field strengths are those obtained from the multi-wavelength model (Table~\ref{tab:model-parameters} and Figure~\ref{fig:SED}). 
The injected particles undergo radiative losses in the magnetic and soft radiation fields, both assumed to be uniform and isotropic. 
This leads to a change in the electron spectrum at each $t_{\mathrm{step}}$. 
Particles move along the $z$ axis by the combination of diffusion and advection. 
Diffusion parallel to the jet direction is included as a random Gaussian smearing with scale $\sqrt{2Dt_{step}}$, where $D$ is the spatially homogeneous diffusion coefficient. Diffusion is parameterised as $D=D_{100}\left(\frac{E}{100\mathrm{TeV}}\right)^{1/3}$, where $=D_{100}$ is the value of the diffusion coefficient at 100~TeV. 
We do not consider diffusion orthogonal to the jet axis. 
The assumed energy dependence of the diffusion coefficient follows Kolmogorov scaling $D\propto E^{1/3}$, consistent with inferences for Galactic cosmic-ray transport~\cite{Strong2007}. 
The advection flow is described by a spatially-dependent velocity $v_{\mathrm{jet}}(z)$. 
Diffusion moves particles in both the increasing and decreasing $z$ directions but advection only transports particles towards increasing values of $z$, away from the acceleration site. 
Figure~\ref{fig:sketch} shows a schematic diagram of this model. 

When the time reaches $t_{\mathrm{inj}}$ we output the a two-dimensional distribution of the number of electrons as a function of their energy and position along the $z$ direction. 
Electrons injected earlier in the simulation time have suffered more cooling losses (Figure~\ref{fig:cooling}) and on average are also farther away from the injection location than freshly injected electrons, due to advection. 
With the computed electron distribution as input, the resulting radiation spectrum due to synchrotron and IC emission was calculated using gamera, including Klein-Nishina corrections to the IC spectrum. 
Simulated emission profiles for selected photon energies and positions along the $z$ direction were be determined. 
We fitted the predicted flux from the model to the measured gamma-ray profiles with two free parameters: the velocity at the base of the outer jets $v_0 = v_{\mathrm{jet}}(0)$ and the diffusion coefficient $D_{100}$. The results are shown in Figure~\ref{fig:profile}.

The only deviation from the observed trend of lower energy gamma rays reaching peak surface brightness further away from the central binary is the lowest-energy bin on the eastern side. 
In this case, sub-threshold excess emission is only detected outside of the \xray jets region and close to the outer jet base. 
However, the 1D model prediction for the spatial distribution of the gamma-ray emission is still compatible with the flux measured at that location within the 1$\sigma$ uncertainty.

\subsection*{Velocity profiles of the jets}

\begin{figure}[]
	\centering
	\includegraphics[width=0.5\textwidth]{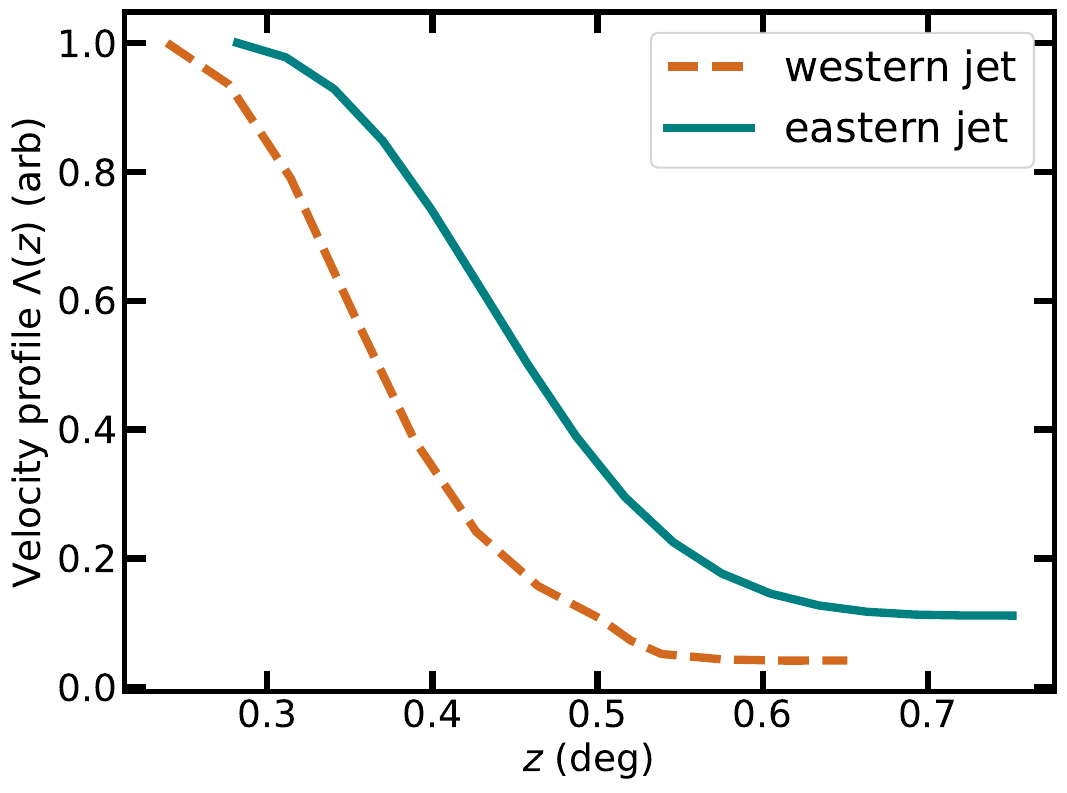}
	
	\caption{\textbf{Calculated deceleration profile for each of the jets. }Shape of the deceleration profile $\Lambda(z)$ as a function of distance from the central binary $z$. The profile is derived from the \xray data~\cite{Brinkmann1996,Safi-Harb1997} for the western (red line) and eastern (blue line) jets. The curves have been smoothed with a Gaussian kernel with a width 0.07\degree\xspace, approximately the \hess PSF.}
	\label{fig:input_velocity_profile}
\end{figure}
The jets are expanding into slightly different ambient conditions: the western jet advances toward the direction of the Galactic plane [\hspace{1sp}\cite{Goodall2011a}, their figure~1], leading to an increase in the ambient density. 
The termination regions of the eastern and western jets lie at a vertical distance from the Galactic plane of approximately 320 and 120~pc, respectively. 
This probably explains the observed different jet length on each side~\cite{Goodall2011a}. 
The flow velocity along the jet, $v_{\mathrm{jet}}(z)$ was characterised as the product of the value of the velocity at the base, $v_0$ and the normalised shape of the velocity profile $\Lambda(z)$, such that $v_{\mathrm{jet}}(z) = v_0 \cdot \Lambda(z)$. 
We constructed the profile $\Lambda(z)$ assuming the jet is cylindrically symmetric and approximately incompressible, meaning that as the jet decelerates, its cross-section increases to preserve the density. Assuming the jet axis to be perpendicular to the line of sight [the measured angle is $\approx 80$\degree\cite{Roberts2010}], the width of the jet can be estimated using the \xray data~\cite{Brinkmann1996,Safi-Harb1997}. 
The absolute value of the jet width is not required for the model, only its evolution as a function of distance from the base of the outer jet. 
The velocity profile $\Lambda(z)$ was computed as the inverse square of the width and normalised such that $\Lambda(0)=1$. 
The resulting velocity profiles for each of the jets are shown in Figure~\ref{fig:input_velocity_profile}, smoothed with the \hess PSF. 
Variations of this profile on scales smaller than the H.E.S.S. resolution will not affect our results. 
The best-fitting value of $v_0=(0.083\pm0.026_{\mathrm{stat}})c$ results in a velocity profile $\Lambda(z)$ which is consistent consistent with the upper limit of 0.023$c$ derived around the termination regions furthest from the central binary.

\subsection*{Systematic uncertainties on the velocity}
\begin{figure}
	\centering
	\includegraphics[width=0.97\textwidth]{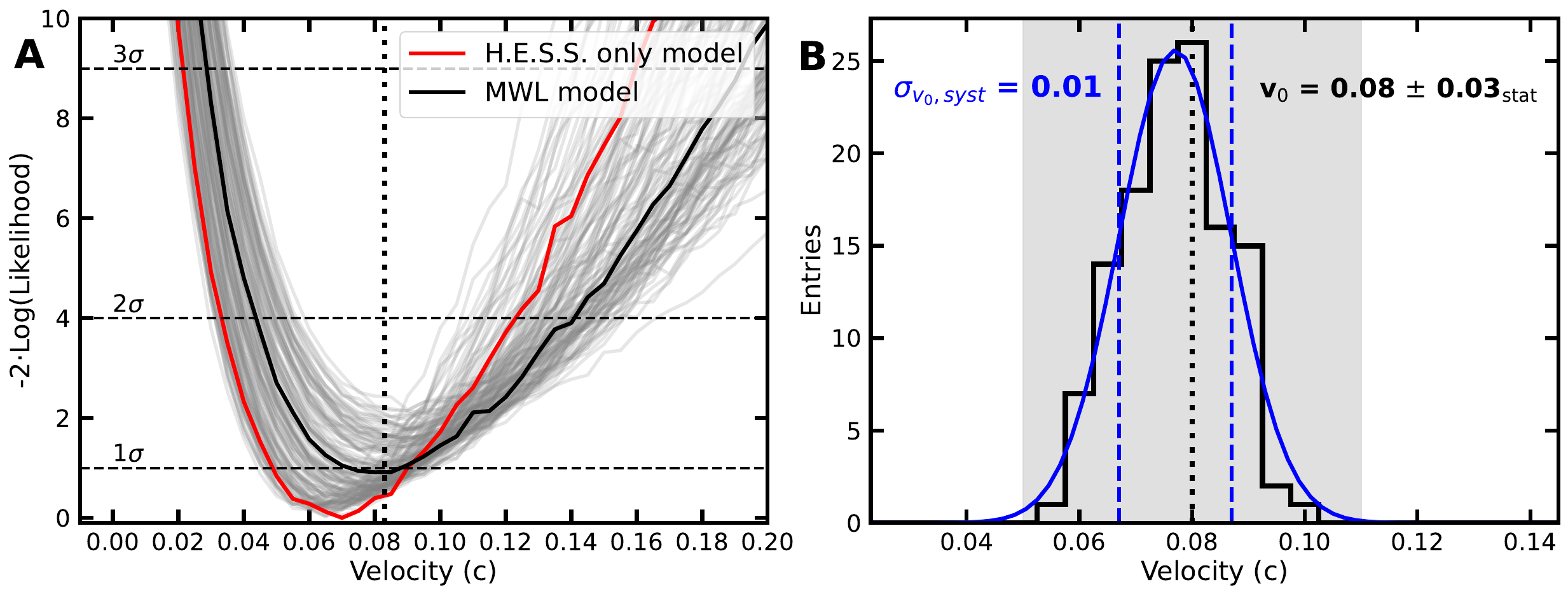}
	\caption{\textbf{Estimation of systematic uncertainties due to the model parameters. }\textbf{A}: Likelihood profiles of the velocity parameter for each of the considered combinations of magnetic field and electron spectral parameters (grey lines). The black solid line shows the likelihood profile corresponding to the parameters obtained from the fit to the multi-wavelength (indicated as MWL) SED. The black dotted line indicates the best-fitting value of the velocity from the multi-wavelength SED. Black dashed lines mark the likelihood values corresponding to 1, 2 and 3 $\sigma$ significance levels. The red solid line shows the best-fitting model when only \hess data is considered. \textbf{B}: Distribution of the v$_0$ values from all the models. The black dotted line and grey band indicate the best-fitting value and statistical uncertainty of the velocity from the multi-wavelength SED. The blue line shows a Gaussian function fitted to the histogram, where the fitted width is indicated by the blue dashed lines. This value is an estimate of the systematic uncertainty $\sigma_{v_0, syst}$. }
	\label{fig:velocity_systematics}
\end{figure}

To assess the impact of the choice of injected electron spectrum parameters and magnetic field on the derived value of $v_0$, we investigated alternative combinations of these parameters. 
We fixed $\Gamma_e$ and $\log_{10}($E$_{\mathrm{cut}})$ to values in the ranges 1.6 to 2.4 and 1.8 to 4, respectively, in steps of 0.05 for both parameters.
The remaining free parameters (magnetic field and normalisation of the electron spectrum) were fitted to the multi-wavelength data (Figure~\ref{fig:SED}). 
All the parameter sets defined by the interior of the 2$\sigma$ likelihood surface in the $\Gamma_e$ and $\log_{10}($E$_{\mathrm{cut}})$ space were selected. 
For each of these electron spectrum parameters, the best-fitting value of $v_0$ was derived by fitting the spatial model prediction to the gamma-ray profiles shown in Figure~\ref{fig:profile} under the assumption of a decelerating jet. 
The diffusion coefficient was fixed, otherwise the fit was not stable for some of the combinations. 
Figure~\ref{fig:velocity_systematics}A shows the likelihood profiles for each of these models. 
This approach only uses the \hess data, so the parameters from the multi-wavelength model are not necessarily the best-fitting combination, although they lie within 1$\sigma$ of the minimum. 
Figure~\ref{fig:velocity_systematics}B shows the distribution of the best-fitting values of $v_0$ for the considered models, compared to the result obtained with the parameters derived from the SED fit (Table~\ref{tab:model-parameters}). 
All the values of $v_0$ derived with different model parameters lie within the statistical uncertainty band, indicating that the choice of parameters does not introduce a large systematic bias. 

\section*{Supplementary Text}

\subsection*{Jet deceleration}
In our model we have assumed a decelerating jet, as motivated by the apparent observed spread of the \xray emission. 
Here we consider alternative possible jet dynamics to test the robustness of our conclusions. 
We explore the case of a jet moving with constant velocity  $\Lambda(z)=1$ for both constant section (no adiabatic losses) and expanding section (with adiabatic losses $\frac{dE}{dt}=-\frac{1}{3}(\vec{\nabla} \cdot \vec{v}_{\mathrm{jet}})E$ (Figure~\ref{fig:cooling}). 
Fitting the value of $v_0$ to the gamma-ray profiles under these assumptions results in $v_0 = (0.045 \pm 0.014_{\mathrm{stat}})c$  and $v_0 = (0.061 \pm 0.013_{\mathrm{stat}})c$ for the cases with and without adiabatic losses, respectively. 
These values of $v_0$ are consistent (within the uncertainty) with the value obtained assuming deceleration, $v_0=(0.083\pm0.026_{\mathrm{stat}})c$.
Any scenario in which $v_0=0$ is disfavoured by more than 5$\sigma$, so regardless of the jet internal structure, advection of the particles is required to explain the observations. 
We cannot distinguish between the different jet propagation scenarios with the available observations.

\subsection*{Diffusion}
The value of $D_{100}$ is fitted at the same time as $v_0$ to the gamma-ray data shown in Figure~\ref{fig:profile}. 
The resulting best-fitting values are $D_{100}=(2.3\pm 1.4) \cdot 10^{28}$~cm$^2$~s$^{-1}$ when assuming a decelerating jet and $D_{100}=(4.7\pm 4.1) \cdot 10^{27}$~cm$^2$~s$^{-1}$ when assuming the velocity to be constant (with constant section). 
Comparing the advection and diffusion lengths within a cooling time $t_{\mathrm{cool}}$ (Figure~\ref{fig:cooling}) we find that $D\ll \frac{t_{\mathrm{cool}}v_0^2}{2}$ at all relevant energies, which indicates that advection is the dominant particle transport process taking place in the outer jets of \ssftt. 
The best-fitting value of $D_{100}$ is approximately an order of magnitude lower than the average Galactic diffusion coefficient~\cite{Strong2007}, which we interpret is due to the stronger magnetic field in \ssftt. There is no a priori reason to expect the diffusion coefficient to match the Galactic value, given that the properties of the medium inside W50 are likely determined by the jets themselves and are unlikely to be similar to the average interstellar medium.

\subsection*{Magnetic field}
The magnetic field strengths $B$ reported in Table~\ref{tab:model-parameters} were derived from a fit to the multi-wavelength SED of the jets, including both the \xray and \gammaray data (Figure~\ref{fig:SED}). 
The energy-dependent morphology provides an alternative way to estimate the average magnetic field in the outer jets that is independent of the \xray observations. 
The dominant processes responsible for the observed morphology are the advection of particles in the jet flow, parameterised by $v_0$ and energy losses due to synchrotron emission, parameterised by $B$. 
We fit the model gamma-ray profiles to the observed data with $v_0$, $B$ and the normalisation $\alpha$ as free parameters, with the rest of the parameters fixed to the values reported in Table~\ref{tab:model-parameters}. 
We do this for both the constant velocity and the decelerating flow case. 
We find that the best-fitting value for the average magnetic field strength in both outer jets is 21.0$\pm$1.7~\textmu G and 20.5$\pm$2.5~\textmu G for the constant and decelerating flows respectively. 
These magnetic field values are consistent with those obtained from the SED, implying that the estimate of the average magnetic field reported in Table~\ref{tab:model-parameters} is consistent with the observed energy-dependent morphology. 
The values of $v_0$ are consistent with those found for the fixed magnetic field, but with larger statistical uncertainties due to the additional free parameter.

\subsection*{Contribution from hadronic processes}
\begin{figure}
	\centering
	\includegraphics[width=0.9\textwidth]{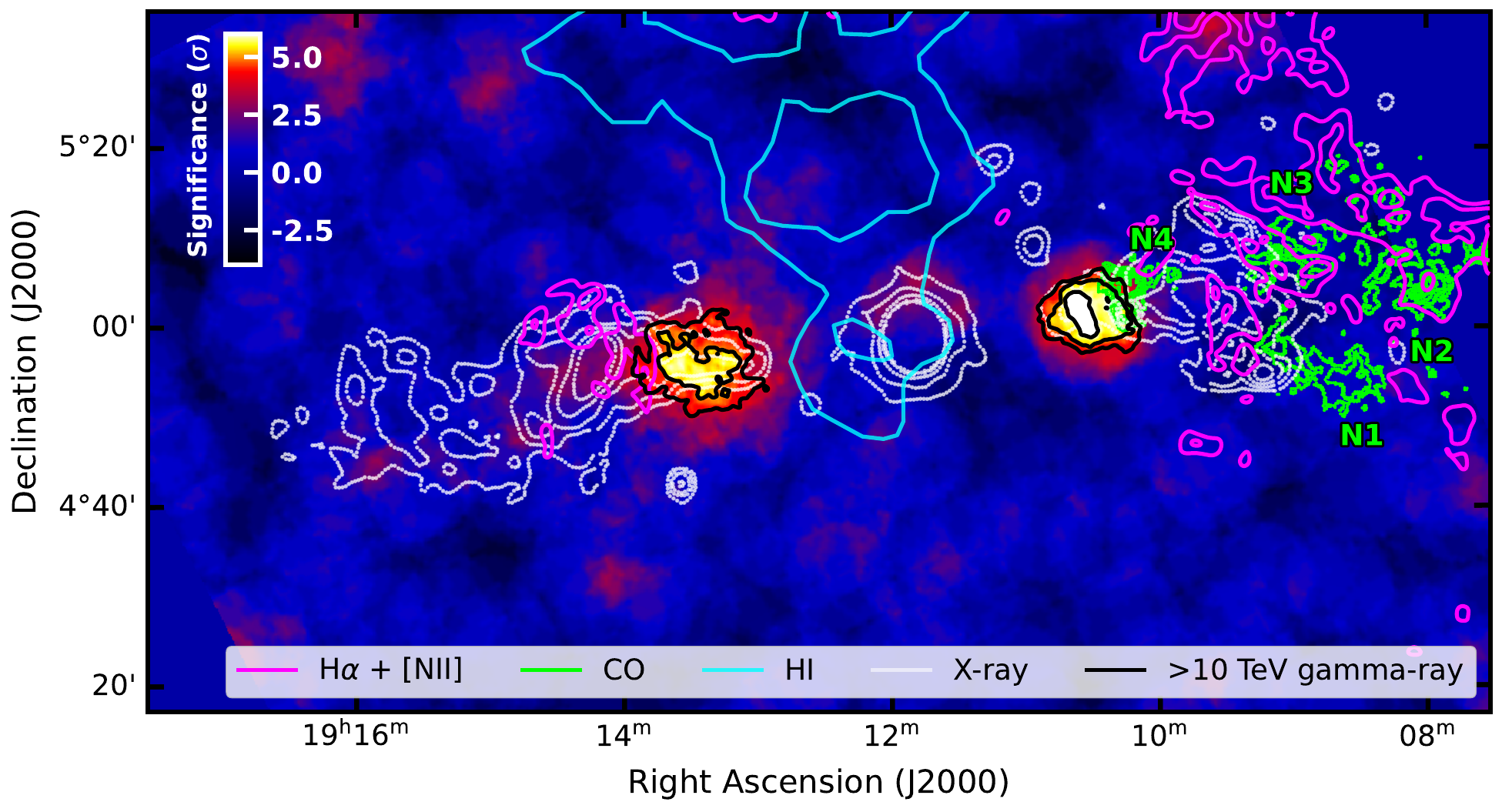}
	\caption{\textbf{Location of possible sites of hadronic interactions.} The \hess significance map above 10~TeV (rotated from the orientation in Figure~\ref{fig:significance-energy}C) compared to observed gas locations, which are possible target material for hadronic interactions. Equatorial coordinates are shown for the J2000 equinox. Black contours indicate significances of 4, 5 and 6$\sigma$ in the \hess map. The pink contour indicates H$\alpha$ + [N$\textsc{ii}$] emission from ionised gas~\cite{Boumis2007}, green corresponds to CO observations revealing four molecular clouds N1 to N4~\cite{Yamamoto2022} and light blue to neutral hydrogen emission from diffuse neutral gas~\cite{Su2018}. The ROSAT \xray contours~\cite{Brinkmann1996} are shown for reference in white. There is no correlation between any of the potential targets and the \hess emission above 10~TeV.}
	\label{fig:hadronic}
\end{figure}

The observed energy-dependent morphology requires the bulk of the gamma-ray emission to be the result of IC emission from electrons. 
The cooling time of protons via hadronic interactions with surrounding gas at high proton energies is nearly independent of energy and inversely proportional to the gas density~\cite{Aharonian2004}. Consequently, protons lose energy slowly for any expected value of gas density in the outer jets (see below) and across all relevant proton energies. In contrast, the electrons cool rapidly at high energies via inverse Compton and synchrotron losses. 
While the presence of accelerated hadrons in the jets cannot be ruled out by our observations, we rule out a large contribution of hadronic emission to the observed \gammaray flux due to the lack of dense target material in the jets region.  
While the ambient density in the \ssftt/W50 complex is unknown, observations~\cite{Su2018} and simulations~\cite{Goodall2011a} have derived values in the 0.1 to 2~cm$^{-3}$ range. 
If we attempt to model the H.E.S.S. observations as proton-proton (p-p) interactions using gamera, we find that for values of the ambient density below 3~cm$^{-3}$, more than 100\% of the available jet power needs to be injected into protons to match the observed flux. 
To reach the observed fluxes while requiring lower fractions ($\sim$10\%) of the available power, p-p interactions would require a denser ($>$20~cm$^{-3}$) medium. 
However, the absence of spatial correlation between the \gammaray emission, especially at the highest energies, and the presence of dense target material (Figure~\ref{fig:hadronic}) provide an additional indication that the majority of the observed emission is due to relativistic electrons.

While the observed gamma-ray emission is dominated by the inverse Compton emission of electrons, we expect protons or heavier nuclei to be accelerated to similar or greater energies at the same acceleration site as the TeV emitting electrons.
Radio observations have detected several molecular clouds in the vicinity of the system~\cite{Band1989}, some aligned with the \ssftt jet axis~\cite{Su2018}, mostly in the western extremity of W50~\cite{Liu2020,Yamamoto2022, Sakemi2021}. 
Their kinematics indicate a possible connection with \ssftt~\cite{Liu2020, Yamamoto2022}. 
However, there are no distance estimates to these clouds, so their position with respect to the jets remains unknown. 
We do no observe gamma-ray emission from any of these clouds.

\subsection*{\ssftt as a cosmic-ray source}
Microquasars have been proposed as candidate sources of the Galactic cosmic-rays~\cite{Heinz2002,Fender2005,Cooper2020,Escobar2022}. 
The hard injection spectrum and high maximum energy inferred from our and \xray observations are consistent with this proposal, though \ssftt is unlikely to contribute to the local cosmic-ray flux at Earth.
Given an assumed upper limit on the system age of $t=30\,000$ years and adopting the average Galactic diffusion coefficient $D_{\mathrm{gal}}=10^{28}\left(\frac{E}{1\mathrm{GeV}}\right)^{1/3}$~cm$^2$~s$^{-1}$~\cite{Strong2007}, the distance that cosmic rays can traverse via diffusive propagation is approximately $r=\sqrt{4D_{\mathrm{gal}}t}$. 
Even for proton energies of 1~PeV, this gives $r\approx0.6\cdot \left(\frac{t}{30\,000\text{ yr}}\right)^{1/2}$~kpc, much smaller than even the lowest estimates for the distance to \ssftt of 3.8~kpc~\cite{Arnason2021}. 
To reach that distance, the average Galactic diffusion coefficient would need to be several orders of magnitude larger than predicted, or the system would need to be around 40 times older, which is incompatible both with the GeV measurements (Figure~\ref{fig:age}) and with the highest estimates of the age of W50 of 100\,000~yr.~\cite{BowlerKeppens2018}. 
Although we expect protons and other nuclei to be accelerated in the jets of \ssftt, we conclude that they do not contribute to the cosmic-ray flux measured on Earth.\\

We nevertheless consider the potential contribution of microquasars to the average Galactic cosmic ray population. Taking our derived value of the magnetic field ($B$), and the extent of the jet base in \xray observations~\cite{Safi-Harb2022} as a proxy for the shock width ($R$), we infer from the Hillas limit~\cite{Hillas1984}, a maximum energy $E_{\rm Hillas}$ of 
\begin{equation}
	E_{\rm Hillas} \approx 10 Z \left(\frac{B}{20 \mu \mathrm{G}}\right) \left(\frac{u_1}{0.26 c}\right)\left(\frac{R}{1.6 \mathrm{pc}}\right)\mathrm{PeV},
\end{equation}
where $Z$ is the atomic number. 
While this exceeds the maximum electron energy discussed above (Figure~\ref{fig:cooling}), protons and other nuclei experience less radiative losses. 
Systems similar to \ssftt, should they exist, provide intermittent contributions to the Galactic cosmic-ray budget at a few PeV. 
If similar acceleration occurs in extra-galactic jets on larger scales, they could reach the EeV regime of ultra-high-energy cosmic rays.

\end{document}